\RequirePackage{fix-cm}
\RequirePackage{fixltx2e}
\documentclass[english]{article}
\usepackage[T1]{fontenc}
\usepackage[latin9]{inputenc}
\usepackage{geometry}
\usepackage[active]{srcltx}
\usepackage{color}
\usepackage{babel}
\usepackage{float}
\usepackage{enumitem}
\usepackage{amsmath}
\usepackage{amsthm}
\usepackage{amssymb}
\usepackage{graphicx}
\usepackage{setspace}
\usepackage[authoryear]{natbib}
\usepackage[unicode=true,pdfusetitle,
 bookmarks=true,bookmarksnumbered=true,bookmarksopen=true,bookmarksopenlevel=1,
 breaklinks=true,pdfborder={0 0 1},backref=false,colorlinks=true]
 {hyperref}
\hypersetup{
 linkcolor=blue, citecolor=blue, urlcolor=blue, filecolor=blue,pdfpagelayout=OneColumn, pdfnewwindow=true,pdfstartview=XYZ, plainpages=false, pdfpagelabels, hyperindex=true}

\makeatletter

\providecommand{\tabularnewline}{\\}
\newcommand{\lyxdot}{.}

\numberwithin{equation}{section}
\numberwithin{figure}{section}
\numberwithin{table}{section}

\usepackage{pdflscape}
\usepackage{fancyhdr}
\usepackage{lastpage}
\usepackage{amsmath}
\usepackage{graphicx}
\usepackage{color}
\usepackage{fancybox} 
\usepackage{moreverb} 
\usepackage{listings} 
\usepackage{algorithmic}
\floatname{algorithm}{Algorithm}

\usepackage{ifpdf} 
\ifpdf 

 \IfFileExists{lmodern.sty}{\usepackage{lmodern}}{}

\fi 

\let\myTOC\tableofcontents
\renewcommand\tableofcontents{%
  \pdfbookmark[1]{\contentsname}{}
  \myTOC
}

\def\LyX{\texorpdfstring{%
  L\kern-.1667em\lower.25em\hbox{Y}\kern-.125emX\@}
  {LyX}}

\@addtoreset{footnote}{section}

\usepackage{dsfont}
\usepackage{bbm}

\usepackage{eurosym}
\theoremstyle{remark}

\makeatother

\begin{document}
\title{{\Large{}Skewed target range strategy for multiperiod portfolio optimization}\\
{\Large{}using a two-stage least squares Monte Carlo method}\vspace{1em}
}
\author{Rongju Zhang\thanks{Corresponding author. Email: \protect\href{mailto:rongju.zhang\%40monash.edu}{rongju.zhang@monash.edu}} \thanks{CSIRO Data61, RiskLab Australia}\ \thanks{School of Mathematical Sciences, Monash University, Australia},
Nicolas Langren\'e\footnotemark[2], Yu Tian\footnotemark[3], Zili
Zhu\footnotemark[2], Fima Klebaner\footnotemark[3] and Kais Hamza\footnotemark[3]}
\date{\date{}\vspace{-1em}
}
\maketitle
\begin{center}
First version: August 19, 2016\vspace{-1em}
\par\end{center}

\begin{center}
This revised version: September 10, 2018\vspace{-1em}
\par\end{center}

\begin{center}
Final version: \textit{Journal of Computational Finance}, 2019
\par\end{center}
\begin{abstract}
\noindent In this paper, we propose a novel investment strategy for
portfolio optimization problems. The proposed strategy maximizes the
expected portfolio value bounded within a targeted range, composed
of a conservative lower target representing a need for capital protection
and a desired upper target representing an investment goal. This strategy
favorably shapes the entire probability distribution of returns, as
it simultaneously seeks a desired expected return, cuts off downside
risk and implicitly caps volatility and higher moments. To illustrate
the effectiveness of this investment strategy, we study a multiperiod
portfolio optimization problem with transaction costs and develop
a two-stage regression approach that improves the classical least
squares Monte Carlo (LSMC) algorithm when dealing with difficult payoffs,
such as highly concave, abruptly changing or discontinuous functions.
Our numerical results show substantial improvements over the classical
LSMC algorithm for both the constant relative risk-aversion (CRRA)
utility approach and the proposed skewed target range strategy (STRS).
Our numerical results illustrate the ability of the STRS to contain
the portfolio value within the targeted range. When compared with
the CRRA utility approach, the STRS achieves a similar mean\textendash variance
efficient frontier while delivering a better downside risk\textendash return
trade-off.

\vspace{2em}

\noindent \textbf{Keywords}: target-based portfolio optimization;
alternative performance measure; multiperiod portfolio optimization;
least squares Monte Carlo; two-stage regression\vspace{2em}

\noindent \textbf{JEL Classification}: G11, D81, C63, C34, \textbf{MSC
Classification}: 91G10, 91G80, 91G60
\end{abstract}
\newpage{}

\section{Introduction}

A crucial and long-standing problem in the theory and practice of
portfolio optimization is the choice of an effective and transparent
performance criterion that balances risk and return. In this paper,
we propose a novel portfolio optimization criterion that aims to combine
to some extent the respective strengths of the classical criteria
considered in the literature.

The origin of the literature corresponds to the notion of decision
making under uncertainty. From there, \citet{vonNeumann1944} proposed
the expected utility approach for which the investment preferences
are captured by a utility function. The shortcomings of this approach
include the abstract nature of utility functions, which can make them
impractical, and its omission of several practical aspects of actual
decision making, as identified by \citet{Tversky1992}'s cumulative
prospect theory, see for example \citet{Barberis2012}.

The mean-variance framework of \citet{Markowitz1952}, which uses
variance to measure risk, can well approximate the quadratic utility
case. When asset returns are assumed to be normally distributed, many
other risk measures have been found equivalent to variance (for example,
the equivalence to the first and second-order lower partial moments
has been proved by \citet*{Klebaner2017}), but the mean-variance
framework greatly benefits from its simple quadratic formulation.

Some may argue that variance is an inadequate measure of portfolio
risk as asset returns usually exhibit the so-called leptokurtic property,
meaning that higher moments may need to be incorporated into the optimization.
We refer to \citet{Lai1991} and \citet*{Konno1993} for the skewness
component and \citet{Davis1990} for both skewness and kurtosis. Another
approach to address the issue of non-normality of asset returns is
to use a downside risk measure. The most common downside risk measures
are the lower-partial moments (e.g., semivariance introduced in \citet{Markowitz1959}),
Value at Risk (VaR, \citealt{Longerstaey1996}) and Conditional Value
at Risk (CVaR, \citealt{Rockafellar2000}, a.k.a. expected shortfall).
These measures can replace variance to form a mean-downside risk approach,
see \citet{Harlow1991} for a mean-lower-partial moment framework,
\citet{Alexander2002} for the mean-VaR framework and \citet{Agarwal2004}
for the mean-CVaR framework.

The last main strand of literature corresponds to target-based strategies
that aim to track a prespecified investment target. A popular target-based
strategy is to maximize the probability of achieving a return target,
see \citet{Browne1999a} for a fixed absolute target and \citet{Browne1999b},
\citet{Pham2003}, \citet*{Gaivoronski2005} and \citet*{Morton2006}
for relative benchmark targets. Alternatively, one can minimize the
probability of an undesirable outcome, see for example \citet*{Hata2010},
\citet{Nagai2012} and \citet*{Milevsky2006}. Using an explicitly
specified investment target in portfolio optimization makes it easier
to understand and monitor in practice. However, choosing a suitable
investment target that properly balances risk and return remains a
challenging task.

Building upon these classical investment criteria, we propose in this
paper the so-called \textit{Skewed Target Range Strategy} (STRS),
which maximizes the expected portfolio value bounded within a prespecified
target range, composed of a conservative lower target representing
a need for capital protection and a desired upper target corresponding
to an ideal return level the investor wishes to achieve. Implicitly,
the optimization can be described as maximizing the probability that
the realized return lies within the targeted range and as close to
the upper target as possible.

There are three main motivations behind the proposed STRS. The first
motivation traces back to the primary purpose of an investment objective
function, which is to carve a desirable shape for the probability
distribution of returns. The STRS, seeking a desirable expected return
while chopping off most of the tails of the distribution beyond the
targeted range, restrains the entire return distribution. The second
motivation comes from the difficulty of specifying a single return
target for classical target-based strategies, which cannot simultaneously
serve the pursuit of a desired investment target and downside protection.
The STRS solves this dilemma by using an upper target which accounts
for return-seeking preference, combined with a lower target which
accounts for loss-aversion preference. Finally, performance criteria
such as utility functions depending on abstract parameters with unforeseeable
practical effects are unlikely to be adopted by investors. Our proposition
of two explicit targets labeled in terms of returns, with intuitive
purposes (capital protection for the lower target and desired investment
return for the upper target), serves as a more practical investment
criterion.

To test the effectiveness of the proposed STRS (formulated in Section
\ref{sec:objective}), we study a multiperiod portfolio optimization
problem with proportional transaction costs. To do so, we modify the
classical Least Squares Monte Carlo (LSMC) algorithm to use a two-stage
regression technique, which makes the problem of approximating the
abrupt STRS objective function (equation \eqref{eq:objective}) as
easy as approximating a linear function. The LSMC literature and the
details of the proposed two-stage LSMC method are further discussed
in Section \ref{sec:LSMC}. We show that this two-stage LSMC method
is numerically more stable than the classical LSMC method for both
the smooth constant relative risk aversion (CRRA) utility approach
and the abrupt STRS. We find that an appropriate level for the lower
target is the initial portfolio value, as it marginally minimizes
the standard deviation and the downside risk of the terminal portfolio
value. Importantly, we show that the STRS criterion does behave as
expected from its design: the portfolio value is well targeted within
the specified range, and the downside risk is robust with respect
to the choice of the upper target. We numerically show that the STRS
achieves a similar mean-variance efficient frontier while delivering
a better downside risk-return trade-off when compared to the CRRA
utility optimization approach. We also provide two simple extensions
of the STRS, described in Section \ref{sec:Extensions}. The first
extension, dubbed \textit{Flat Target Range Strategy} (FTRS), corresponds
with pure probability maximization of achieving a targeted range,
without a further attempt to pursue a higher return. The FTRS is useful
for problems where maintaining solvency is more important than seeking
high returns, for example for long-term pension schemes, retirement
funds and life-cycle management. The second extension, dubbed \textit{Relative
Target Range Strategy} (RTRS), focuses relative returns: it involves
a return target range defined in terms of excess return over a stochastic
benchmark, such as stock market index, interest rate or inflation
rate. All the numerical results are presented in Section \ref{sec:Numerical}.

\section{Skewed Target Range Strategy\label{sec:objective}}

\begin{figure}
\caption{Skewed target range function\label{fig:STRS}}

\smallskip
\centering{}%
\begin{minipage}[t]{0.45\columnwidth}%
\includegraphics[scale=0.55]{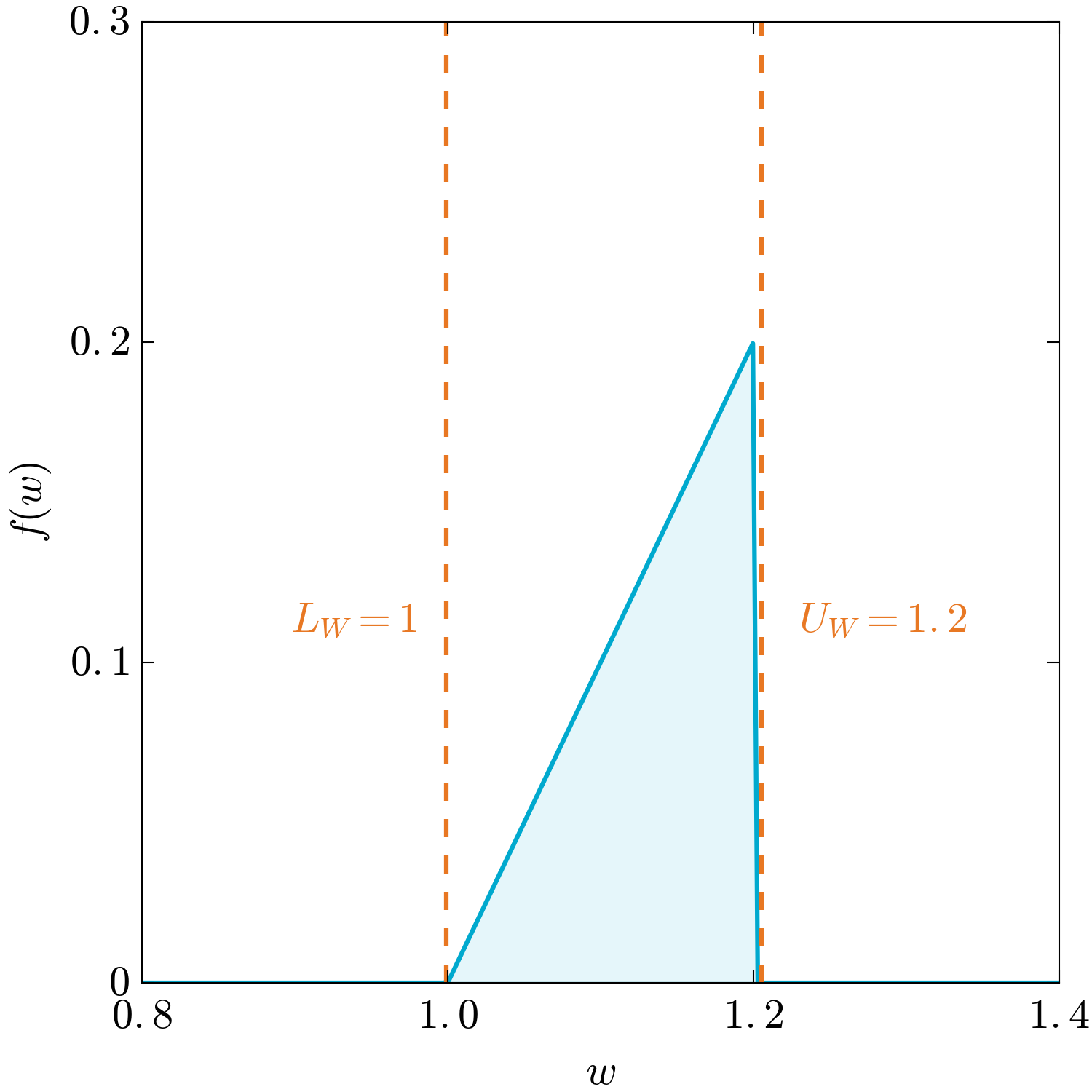}%
\end{minipage}
\end{figure}

In this section, we define the skewed target range strategy (STRS)
for portfolio optimization problems and discuss potential benefits
of this strategy. We consider a portfolio optimization problem with
$d$ risky assets available over a finite time horizon $T$. Let $\boldsymbol{\alpha}_{t}=\{\alpha_{t}^{i}\}_{1\leq i\leq d}$
be the portfolio weight in each risky asset at time $t$, and denote
by $W_{t}$ the portfolio value (or wealth). Assume that the investor
aims to maximize the expectation of some function of the terminal
portfolio value $\mbox{\ensuremath{\mathbb{E}}}\left[f(W_{T})\right]$.
Then, the objective function simply reads 
\begin{equation}
\sup_{\boldsymbol{\alpha}}\mbox{\ensuremath{\mathbb{E}}}\left[f\left(W_{T}\right)\right],\label{eq:objective}
\end{equation}
where the investment preference is characterized by the function $f\left(\cdot\right)$
. In this paper, we propose the following parametric shape: 
\begin{equation}
f(w)=(w-L_{{\scriptscriptstyle \!W}})\mathbbm{1}\{L_{\!{\scriptscriptstyle W}}\leq w\leq U_{{\scriptscriptstyle \!W}}\},\label{eq:f}
\end{equation}
where $L_{{\scriptscriptstyle \!W}}\in\mathbb{R}$ represents a conservative
lower target, $U_{{\scriptscriptstyle \!W}}\in\mathbb{R}$ represents
a desired upper target, and the indicator function $\mathbbm{1}\{L_{{\scriptscriptstyle \!W}}\leq w\leq U_{{\scriptscriptstyle \!W}}\}$
returns one if $L_{{\scriptscriptstyle \!W}}\leq w\leq U_{{\scriptscriptstyle \!W}}$
and returns zero otherwise. We refer to the shape \eqref{eq:f} and
the corresponding objective \eqref{eq:objective} as the STRS. Throughout
this paper, we normalize the portfolio value $W$ and the bounds $[L_{{\scriptscriptstyle \!W}},U_{{\scriptscriptstyle \!W}}]$
by the initial portfolio value $W_{0}$. Indeed, the formula \eqref{eq:f}
shows that $f(w;L_{{\scriptscriptstyle \!W}},U_{{\scriptscriptstyle \!W}})=W_{0}\times f(\frac{w}{W_{0}};\frac{L_{{\scriptscriptstyle \!W}}}{W_{0}},\frac{U_{{\scriptscriptstyle \!W}}}{W_{0}})$,
so we can assume without loss of generality that $W_{0}=1$ and set
the bounds $L_{{\scriptscriptstyle \!W}}$ and $U_{{\scriptscriptstyle \!W}}$
in the vicinity of $1$. Figure \ref{fig:STRS} shows an example of
equation \eqref{eq:f} with $L_{{\scriptscriptstyle \!W}}=1.0$ and
$U_{{\scriptscriptstyle \!W}}=1.2$.

From equation \eqref{eq:f}, one can see that the objective is to
maximize the expected terminal portfolio value within the interval
$[L_{{\scriptscriptstyle \!W}},U_{{\scriptscriptstyle \!W}}]$, while
the values outside this interval are penalized down to zero. This
strategy implicitly combines two objectives: maximizing the expected
terminal portfolio value and maximizing the probability that the terminal
portfolio value lies within the chosen target range $[L_{{\scriptscriptstyle \!W}},U_{{\scriptscriptstyle \!W}}]$.

On the left side of the skewed shape in equation \eqref{eq:f}, the
function is convex at the lower target $L_{{\scriptscriptstyle \!W}}$.
This is consistent with the notion from \citet{Tversky1992}'s cumulative
prospect theory that investors tend to be risk-seeking when losing
money. By contrast, on the right side of the skewed shape, the function
is discontinuous and jumps down to zero at the upper target $U_{{\scriptscriptstyle \!W}}$.
This is the distinctive feature of the STRS compared to classical
utility functions as well as cumulative prospect theory. In particular,
the foregoing of the upside potential beyond the upper target $U_{{\scriptscriptstyle \!W}}$
seems to conflict with the non-satiation axiom that people prefer
more to less. The following explains the importance of this upper
threshold.

Everything else being equal (ceteris paribus assumption), one would
expect people to prefer more to less. This axiom in the context of
dynamic stochastic portfolio optimization can be interpreted as follows:
the downside risk being fixed (the left tail of the return distribution),
investors would prefer higher upside potential (a longer right tail
of the return distribution). However, \textit{\emph{after extensive
numerical experiments, we came to the conclusion that }}non-decreasing
utility functions are unable to decouple upside potential from downside
risk. Indeed, pursuing higher upside potential leads to riskier portfolio
decisions, which may result in a return distribution with a large
right tail (gains) as well as a large left tail (losses). As the ceteris
paribus assumption does not apply in this stochastic context, one
cannot rule our the existence of a satiation level. Such a level is
determined by the investor's preference with respect to risk and return.

As upside potential and downside risk are naturally intertwined, the
proposed upper target is able to curtail downside risk by addressing
its main cause - namely the pursuit of excessive upside potential.
As a result, the realized returns can be well contained within the
targeted range with a high degree of confidence, which in several
contexts is more important than allowing for the possibility of rare
windfall returns at the cost of higher downside risk. 

\section{Multiperiod Portfolio Optimization\label{sec:LSMC}}

In this section, we consider a multiperiod portfolio optimization
problem and formulate it as a discrete-time dynamic programming problem,
for which we develop a two-stage LSMC method to solve it. The LSMC
algorithm, originally developed by \citet{Carriere1996}, \citet{Longstaff2001}
and \citet{Tsitsiklis01} for pricing American options, has been extended
to solve dynamic portfolio optimization problems by several researchers.
\citet*{Brandt2005} consider a CRRA utility function and determine
a semi-closed-form solution by solving the first order condition of
the Taylor series expansion of the value function. \citet{Cong2016}
and \citet{Cong2016b} consider a target-based mean-variance objective
function and use a suboptimal strategy to perform the forward simulation
of control variables which are iteratively updated in the backward
recursive programming. Later, \citet{Cong2017} combine \citet{Jain2015}'s
stochastic bundling technique with \citet{Brandt2005}'s method. \citet*{Zhang2018}
consider a CRRA utility function and adopt \citet*{Kharroubi2014}'s
control randomization technique for a portfolio optimization problem
with switching costs including transaction costs, liquidity costs
and market impact.

The aforementioned works solve problems with a continuous payoff function
for which the classical LSMC method can be very effective. By contrast,
highly nonlinear, abruptly changing or discontinuous payoffs can be
more difficult to handle for the LSMC algorithm (\citet{Zhang2018},
\citet*{Balata18}, \citet*{Andreasson2018}). The STRS \eqref{eq:f},
with its abrupt drop at the upper bound $U_{{\scriptscriptstyle \!W}}$,
is such a difficult function. In addition, as the terminal wealth
outside the targeted range are truncated to zero in the value function,
a direct regression on these zeros would forego the original information
from the wealth variable. In this section, we propose a two-stage
LSMC method to overcome these issues.

\subsection{Dynamic programming}

Denote by $R^{f}$ the cumulative return of the risk-free asset over
one single period. Denote by $\mathbf{R}_{t}=\left\{ R_{t}^{i}\right\} _{1\leq i\leq d}$
the excess returns of the risky assets over the risk-free rate and
denote by $\mathbf{Z}_{t}$ the vector of return predictors. The optimization
problem in equation \eqref{eq:objective} can be formulated as a stochastic
control problem with exogenous state variables $\mathbf{Z}_{t}$ and
one endogenous state variable $W_{t}$. Let $\mathcal{A}\subseteq\mathbb{R}^{d}$
be the set of admissible portfolio weights. The value function in
equation \eqref{eq:objective} can now be rewritten as 
\begin{eqnarray}
v_{t}(z,w) & := & \sup_{\left\{ \boldsymbol{\alpha}_{\tau}\in\mathcal{A}\right\} _{t\leq\tau\leq T}}\mbox{\ensuremath{\mathbb{E}}}\left[f\left(W_{T}\right)\left|\mathbf{Z}_{t}=z,W_{t}=w\right.\right].
\end{eqnarray}
Consider an equidistant discretization of the investment horizon $[0,T]$,
denoted by $0=t_{0}<\cdots<t_{N}=T$. The wealth process evolves as
\begin{eqnarray}
W_{t_{n+1}} & = & W_{t_{n}}\left(R^{f}+\boldsymbol{\alpha}_{t_{n}}\cdot\mathbf{R}_{t_{n+1}}\right),\label{eq:wealth}
\end{eqnarray}
and the value function satisfies the following dynamic programming
principle 
\begin{eqnarray}
v_{t_{N}}\left(z,w\right) & = & f(w),\nonumber \\
v_{t_{n}}\left(z,w\right) & = & \sup_{\boldsymbol{\alpha}_{t_{n}}\in\mathcal{A}}\mbox{\ensuremath{\mathbb{E}}}\left[v_{t_{n+1}}\left(\mathbf{Z}_{t_{n+1}},W_{t_{n+1}}\right)\left|\mathbf{Z}_{t_{n}}=z,W_{t_{n}}=w\right.\right],\label{eq:dp}
\end{eqnarray}
where $f(w)=(w-L_{{\scriptscriptstyle \!W}})\mathbbm{1}\{L_{\!{\scriptscriptstyle W}}\leq w\leq U_{{\scriptscriptstyle \!W}}\}$.

\subsection{Classical least squares Monte Carlo }

The first part of the LSMC algorithm is the forward simulation of
all the stochastic state variables. Let $M$ denote the number of
Monte Carlo simulations. The return predictors $\{\mathbf{Z}_{t_{n}}^{m}\}_{0\leq n\leq N}^{1\leq m\leq M}$
and the asset excess returns $\{\mathbf{R}_{t_{n}}^{m}\}_{0\leq n\leq N}^{1\leq m\leq M}$
are generated through some predetermined return dynamics. By contrast,
the wealth process is an endogenous state variable depending on the
realization of the portfolio weights. We follow the control randomization
approach of \citet{Kharroubi2014}: we randomly generate uniform portfolio
weights $\{\tilde{\boldsymbol{\alpha}}_{t_{n}}^{m}\}_{0\leq n\leq N}^{1\leq m\leq M}$,
and then compute the corresponding portfolio values $\{\tilde{W}_{t_{n}}^{m}\}_{0\leq n\leq N}^{1\leq m\leq M}$
according to equation \eqref{eq:wealth}.

The second part of the LSMC algorithm uses a discretization procedure.
We discretize the control space as $\mathcal{A}^{\text{d}}=\{\mathbf{a}_{1},...,\mathbf{a}_{J}\}$.
We define the continuation value function $\text{CV}{}_{t_{n}}^{j}$
as the expectation of the subsequent value function conditional on
making the decision $\boldsymbol{\alpha}_{t_{n}}=\mathbf{a}_{j}\in\mathcal{A}^{\text{d}}$,
i.e., 
\begin{eqnarray}
\text{CV}{}_{t_{n}}^{j}\left(z,w\right) & := & \mathbb{E}\left[\left.v_{t_{n+1}}\left(\mathbf{Z}_{t_{n+1}},W_{t_{n+1}}\right)\right|\mathbf{Z}_{t_{n}}=z,W_{t_{n}}=w,\boldsymbol{\alpha}_{t_{n}}=\mathbf{a}_{j}\right].\label{eq:cv}
\end{eqnarray}

Therefore, the value function can be approximated by
\[
v_{t_{n}}(z,w)=\sup_{\boldsymbol{\alpha}_{t_{n}}\in\mathcal{A}}\mathbb{E}\left[\left.v_{t_{n+1}}\left(\mathbf{Z}_{t_{n+1}},W_{t_{n+1}}\right)\right|\mathbf{Z}_{t_{n}}=z,W_{t_{n}}=w\right]\approx\max_{\mathbf{a}_{j}\in\mathcal{A}^{\text{d}}}\text{CV}{}_{t_{n}}^{j}\left(z,w\right).
\]
To compute this value function, we proceed by backward dynamic programming.
At time $t_{N}$, the value function is equal to $\hat{v}_{t_{N}}(z,w)=(w-L_{{\scriptscriptstyle \!W}})\mathbbm{1}\{L_{{\scriptscriptstyle \!W}}\leq w\leq U_{{\scriptscriptstyle \!W}}\}$.
At time $t_{n}$, assume that the continuation value functions $\{\hat{\text{CV}}{}_{t_{n'}}^{j}(z,w)\}_{n+1\leq n'\leq N-1}^{1\leq j\leq J}$
have been estimated. We evaluate the continuation value function at
the current time $\text{CV}_{t_{n}}^{j}$ for each decision $\mathbf{a}_{j}\in\mathcal{A}^{\text{d}}$.
We then reset the portfolio weights $\{\boldsymbol{\alpha}_{t_{n}}^{m}\}_{1\leq m\leq M}$
to $\mathbf{a}_{j}$, and recompute the endogenous wealth from $t_{n}$
to $t_{N}$: 
\begin{eqnarray}
\hat{W}_{t_{n+1}}^{m,(n,j)} & = & \tilde{W}_{t_{n}}^{m}\left(R^{f}+\mathbf{a}_{j}\cdot\mathbf{R}_{t_{n+1}}^{m}\right)\nonumber \\
\hat{W}_{t_{n+2}}^{m,(n,j)} & = & \hat{W}_{t_{n+1}}^{m,(n,j)}\left(R^{f}+\arg\max_{\mathbf{a}_{l}\in\mathcal{A}^{\text{d}}}\left\{ \hat{\text{CV}}{}_{t_{n+1}}^{l}\left(\mathbf{Z}_{t_{n+1}}^{m},\hat{W}_{t_{n+1}}^{m,(n,j)}\right)\right\} \cdot\mathbf{R}_{t_{n+2}}^{m}\right)\nonumber \\
 & \vdots\nonumber \\
\hat{W}_{t_{N}}^{m,(n,j)} & = & \hat{W}_{t_{N-1}}^{m,(n,j)}\left(R^{f}+\arg\max_{\mathbf{a}_{l}\in\mathcal{A}^{\text{d}}}\left\{ \hat{\text{CV}}{}_{t_{N-1}}^{l}\left(\mathbf{Z}_{t_{N-1}}^{m},\hat{W}_{t_{N-1}}^{m,(n,j)}\right)\right\} \cdot\mathbf{R}_{t_{N}}^{m}\right).\label{eq:wealth_hat-evolution}
\end{eqnarray}
where $\hat{W}_{t_{n'}}^{m,(n,j)}:=\left.\hat{W}_{t_{n'}}^{m}\right|_{W_{t_{n}}^{m}=\tilde{W}_{t_{n}}^{m},\boldsymbol{\alpha}_{t_{n}}=\mathbf{a}_{j}}$,
$n'=n,\ldots,N$ is the recomputed wealth from $t_{n}$ to $t_{N}$,
using the portfolio weights $\mathbf{a}_{j}$ at time $t_{n}$ and
the estimated optimal portfolio weights at times $t_{n+1},\ldots,t_{N-1}$.

To approximate the continuation value function $\text{CV}{}_{t_{n}}^{j}(z,w)$,
the classical LSMC algorithm regresses the payoffs $\{f(\hat{W}_{t_{N}}^{m,(n,j)})\}_{1\leq m\leq M}$
on $\{\psi_{k}(\mathbf{Z}_{t_{n}}^{m},\tilde{W}_{t_{n}}^{m})\}_{1\leq m\leq M}^{1\leq k\leq K}$,
where $\{\psi_{k}(z,w)\}_{1\leq k\leq K}$ is the vector of basis
functions of the state variables. However, the major difficulty here
lies in the abrupt upper bound $U_{{\scriptscriptstyle \!W}}$, which
can cause large numerical errors in the regression according to our
numerical exploration.

As $f$ censors the values of $\hat{W}_{t_{N}}^{m,(n,j)}$ outside
the targeted range $[L_{{\scriptscriptstyle \!W}},U_{{\scriptscriptstyle \!W}}]$,
our regression problem looks similar to a censored regression problem,
for which a common estimation approach is maximum likelihood estimation
(MLE). However, the main difference between our problem and a censored
regression problem is that we have access to both the censored samples
$\{f(\hat{W}_{t_{N}}^{m,(n,j)})\}_{1\leq m\leq M}$ and the uncensored
samples $\{\hat{W}_{t_{N}}^{m,(n,j)}\}_{1\leq m\leq M}$. Thus, MLE
would ignore the information of the uncensored values $\hat{W}_{t_{N}}^{m,(n,j)}$
which are also observable in this estimation problem. The availability
of this extra piece of information motivates us to propose a two-stage
regression that takes advantages of this information. We now describe
this technique in detail.

\subsection{Two-stage least squares Monte Carlo \label{subsec:TLSMC}}

This two-stage regression works as follows: 
\begin{enumerate}
\item Instead of regressing the payoffs $\{f(\hat{W}_{t_{N}}^{m,(n,j)})\}_{1\leq m\leq M}$,
we regress the wealth $\{\hat{W}_{t_{N}}^{m,(n,j)}\}_{1\leq m\leq M}$
on $\{\psi_{k}(\mathbf{Z}_{t_{n}}^{m},\tilde{W}_{t_{n}}^{m})\}_{1\leq m\leq M}^{1\leq k\leq K}$
to obtain 
\begin{eqnarray}
\left\{ \hat{\beta}_{k,t_{n}}^{j}\right\} {}_{1\leq k\leq K} & = & {\displaystyle \arg\min_{\beta\in\mathbb{R}^{K}}}\sum_{m=1}^{M}\left(\sum_{k=1}^{K}\beta_{k}\psi_{k}\left(\mathbf{Z}_{t_{n}}^{m},\tilde{W}_{t_{n}}^{m}\right)-\hat{W}_{t_{N}}^{m,(n,j)}\right)^{2},\nonumber \\
\hat{\sigma}_{t_{n}}^{j} & = & \sqrt{\frac{1}{M-K}\sum_{m=1}^{M}\left(\hat{W}_{t_{N}}^{m,(n,j)}-\sum_{k=1}^{K}\hat{\beta}_{k,t_{n}}^{j}\psi_{k}\left(\mathbf{Z}_{t_{n}}^{m},\tilde{W}_{t_{n}}^{m}\right)\right)^{2}}.\label{eq:ls}
\end{eqnarray}
As a result, the terminal wealth can be modeled as 
\begin{eqnarray}
\hat{W}_{t_{N}}^{(n,j)}=\hat{\mu}_{t_{n}}^{j}\left(z,w\right)+\hat{\sigma}_{t_{n}}^{j}\varepsilon, &  & \hat{\mu}_{t_{n}}^{j}\left(z,w\right):=\sum_{k=1}^{K}\hat{\beta}_{k,t_{n}}^{j}\psi_{k}\left(z,w\right),\label{eq:linear-wealth}
\end{eqnarray}
where $\varepsilon$ is the regression residual, which for demonstrative
purposes we assume Gaussian. (Remark that an assumption for the distribution
of the residuals is also required by MLE.) Let $\phi(x)=\frac{1}{\sqrt{2\pi}}\exp(\frac{x^{2}}{2})$
represent the standard normal probability density function, and $\Phi(x)=\int_{-\infty}^{x}\phi(x)dx$
represent the standard normal cumulative distribution function. 
\item Plug equation \eqref{eq:linear-wealth} into the continuation value
formula \eqref{eq:cv} to obtain a closed-form estimate. By combining
equations \eqref{eq:cv}, \eqref{eq:wealth_hat-evolution}, \eqref{eq:ls}
and \eqref{eq:linear-wealth}, we obtain the following closed-form
estimate of the continuation value function for each $\mathbf{a}_{j}\in\mathcal{A}^{\text{d}}$
at time $t_{n}$: 
\begin{eqnarray}
\hat{\text{CV}}{}_{t_{n}}^{j}\left(z,w\right) & = & \mathbb{E}\left[\left.\left(W_{t_{N}}-L_{\!{\scriptscriptstyle W}}\right)\mathbbm{1}\left\{ L_{{\scriptscriptstyle \!W}}\leq W_{t_{N}}\leq U_{{\scriptscriptstyle \!W}}\right\} \right|\mathbf{Z}_{t_{n}}=z,W_{t_{n}}=w,\boldsymbol{\alpha}_{t_{n}}=\mathbf{a}_{j}\right]\nonumber \\
 & = & \mathbb{E}_{\varepsilon}\left[\left(\hat{\mu}_{t_{n}}^{j}\left(z,w\right)+\hat{\sigma}_{t_{n}}^{j}\varepsilon-L_{\!{\scriptscriptstyle W}}\right)\times\mathbbm{1}\left\{ L_{{\scriptscriptstyle \!W}}\leq\hat{\mu}_{t_{n}}^{j}\left(z,w\right)+\hat{\sigma}_{t_{n}}^{j}\varepsilon\leq U_{{\scriptscriptstyle \!W}}\right\} \right]\nonumber \\
 & = & \left(\hat{\mu}_{t_{n}}^{j}\left(z,w\right)-L_{\!{\scriptscriptstyle W}}\right)\mathbb{E}_{\varepsilon}\left[\mathbbm{1}\left\{ \frac{L_{{\scriptscriptstyle \!W}}-\hat{\mu}_{t_{n}}^{j}\left(z,w\right)}{\hat{\sigma}_{t_{n}}^{j}}\leq\varepsilon\leq\frac{U_{{\scriptscriptstyle \!W}}-\hat{\mu}_{t_{n}}^{j}\left(z,w\right)}{\hat{\sigma}_{t_{n}}^{j}}\right\} \right]\nonumber \\
 &  & +\hat{\sigma}_{t_{n}}^{j}\mathbb{E}_{\varepsilon}\left[\varepsilon\mathbbm{1}\left\{ \frac{L_{{\scriptscriptstyle \!W}}-\hat{\mu}_{t_{n}}^{j}\left(z,w\right)}{\hat{\sigma}_{t_{n}}^{j}}\leq\varepsilon\leq\frac{U_{{\scriptscriptstyle \!W}}-\hat{\mu}_{t_{n}}^{j}\left(z,w\right)}{\hat{\sigma}_{t_{n}}^{j}}\right\} \right]\nonumber \\
 & = & \left(\hat{\mu}_{t_{n}}^{j}\left(z,w\right)-L_{\!{\scriptscriptstyle W}}\right)\left(\Phi\left(\frac{U_{{\scriptscriptstyle \!W}}-\hat{\mu}_{t_{n}}^{j}\left(z,w\right)}{\hat{\sigma}_{t_{n}}^{j}}\right)-\Phi\left(\frac{L_{{\scriptscriptstyle \!W}}-\hat{\mu}_{t_{n}}^{j}\left(z,w\right)}{\hat{\sigma}_{t_{n}}^{j}}\right)\right)\nonumber \\
 &  & -\hat{\sigma}_{t_{n}}^{j}\left(\phi\left(\frac{U_{{\scriptscriptstyle \!W}}-\hat{\mu}_{t_{n}}^{j}\left(z,w\right)}{\hat{\sigma}_{t_{n}}^{j}}\right)-\phi\left(\frac{L_{{\scriptscriptstyle \!W}}-\hat{\mu}_{t_{n}}^{j}\left(z,w\right)}{\hat{\sigma}_{t_{n}}^{j}}\right)\right),\label{eq:cv-STRS}
\end{eqnarray}
where the last equality is obtained by direct integration. 
\item The mappings $\hat{\boldsymbol{\alpha}}_{t_{n}}:(z,w)\mapsto\hat{\boldsymbol{\alpha}}_{t_{n}}(z,w)$
and $\hat{v}_{t_{n}}:(z,w)\mapsto\hat{v}_{t_{n}}(z,w)$ are estimated
by 
\begin{eqnarray}
\hat{\boldsymbol{\alpha}}_{t_{n}}\left(z,w\right)=\arg\max_{\mathbf{a}_{j}\in\mathcal{A}^{\text{d}}}\hat{\text{CV}}{}_{t_{n}}^{j}\left(z,w\right) & \text{ and } & \hat{v}_{t_{n}}(z,w)=\max_{\mathbf{a}_{j}\in\mathcal{A}^{\text{d}}}\hat{\text{CV}}{}_{t_{n}}^{j}\left(z,w\right).\label{eq:opt}
\end{eqnarray}
\end{enumerate}
In summary, thanks to the censored linear shape of the skewed target
range function in equation \eqref{eq:f}, the conditional expectations
in the dynamic programming equations \eqref{eq:dp} can be estimated
by the closed-form formula \eqref{eq:cv-STRS}. Due to the linearity
of the regressand $\hat{W}_{t_{N}}^{m,(n,j)}$ in equation \eqref{eq:ls},
this two-stage regression is much more robust and stable than a direct
regression of $f(\hat{W}_{t_{N}}^{m,(n,j)})$. Subsection \ref{subsec:CRRA-utility}
describes a similar closed-form conditional value for the CRRA utility
approach, and Subsection \ref{subsec:Model-validation} illustrates
the numerical improvements provided by this two-stage LSMC method.

More generally, the approach proposed here (linear approximation in
\eqref{eq:linear-wealth} + decensored corrections in \eqref{eq:cv-STRS})
can be adapted to the situations where residuals are non-Gaussian:
this would simply modify the correction terms in \eqref{eq:cv-STRS}.
There is no restriction on the choice of the residual distribution,
nor on the estimation methods (empirical distribution, kernel estimation,
mixture normal, etc.). Nevertheless, without loss of generality, it
can be reasonable to assume normality of residuals for low-frequency
trading such as monthly returns with monthly rebalancing considered
in our numerical experiments in Section \ref{sec:Numerical}. In addition,
the properties of the wealth distribution can be well captured by
regressing $\{\hat{W}_{t_{N}}^{m,(n,j)}\}_{1\leq m\leq M}$ on basis
functions of $\{\tilde{W}_{t_{n}}^{m}\}_{1\leq m\leq M}$, yielding
regression residuals close to normal. Based on our numerical experiments,
the residuals are indeed very close to normal. For these reasons and
for demonstration purposes, we henceforth assume normality of residuals
and focus on the analysis of the effects of the new investment objective
\eqref{eq:f}.

\subsection{State-dependent standard deviation\label{subsec:State-dependent-SD}}

An important assumption made in the previous subsection is that $\hat{\sigma}_{t_{n}}^{j}$
only depends on the portfolio decision $\boldsymbol{a}^{j}$, but
not on the state variables $(\mathbf{Z}_{t_{n}},W_{t_{n}})$. This
subsection describes how to improve the standard deviation estimate
to incorporate state variables. Similar to the approximation of $\hat{\mu}_{t_{n}}^{j}(z,w)$,
the state-dependent standard deviation $\hat{\sigma}_{t_{n}}^{j}(z,w)$
can be approximated by the exponential of a linear combination of
basis functions of state variables, $\hat{\sigma}_{t_{n}}^{j}(z,w)=\exp(\sum_{k=1}^{K'}\hat{\eta}_{k,t_{n}}^{j}\psi_{k}\left(z,w\right))$.
The purpose of the exponential transform is to avoid the possibility
of negative standard deviation estimates. Then, the two-stage regression
becomes
\begin{eqnarray*}
\hat{W}_{t_{N}}^{(n,j)} & = & \hat{\mu}_{t_{n}}^{j}\left(z,w\right)+\varepsilon,\\
\varepsilon & \sim & \mathcal{N}\left(0,\hat{\sigma}_{t_{n}}^{j}\left(z,w\right)\right),\\
\hat{\mu}_{t_{n}}^{j}\left(z,w\right) & = & \sum_{k=1}^{K}\hat{\beta}_{k,t_{n}}^{j}\psi_{k}\left(z,w\right),\\
\hat{\sigma}_{t_{n}}^{j}\left(z,w\right) & = & \exp\left(\sum_{k=1}^{K'}\hat{\eta}_{k,t_{n}}^{j}\psi_{k}\left(z,w\right)\right).
\end{eqnarray*}

Note that a standard least squares regression cannot be used to estimate
an unobservable variable such as standard deviation. Instead, we use
MLE. We first perform a least squares regression to approximate the
mean $\hat{\mu}_{t_{n}}^{j}(z,w)$, and then approximate the logarithmic
standard deviation $\log\hat{\sigma}_{t_{n}}^{j}(z,w)$ by maximizing
the following log-likelihood function:
\[
\mathcal{L}\left(\eta\left|\mathbf{Z}_{t_{n}},\tilde{W}_{t_{n}},\hat{W}_{t_{N}}^{(n,j)}\right.\right)=\sum_{m=1}^{M}\left\{ -\sum_{k=1}^{K'}\eta_{k,t_{n}}^{j}\psi_{k}\left(\mathbf{Z}_{t_{n}}^{m},\tilde{W}_{t_{n}}^{m}\right)-\frac{\left(\hat{\varepsilon}^{m}\right)^{2}}{2}\exp\left(-2\sum_{k=1}^{K'}\eta_{k,t_{n}}^{j}\psi_{k}\left(\mathbf{Z}_{t_{n}}^{m},\tilde{W}_{t_{n}}^{m}\right)\right)\right\} ,
\]
where 
\begin{eqnarray*}
\hat{\varepsilon}^{m} & = & \hat{W}_{t_{N}}^{m,(n,j)}-\sum_{k=1}^{K}\hat{\beta}_{k,t_{n}}^{j}\psi_{k}\left(\mathbf{Z}_{t_{n}}^{m},\tilde{W}_{t_{n}}^{m}\right).
\end{eqnarray*}

We use the Broyden\textendash Fletcher\textendash Goldfarb\textendash Shanno
algorithm to perform the maximization of this log-likelihood function.
In Subsection \ref{subsec:Model-validation}, we compare the results
obtained with and without state-dependency in the standard deviation
estimate. 

\subsection{Upper target as stop-profit\label{subsec:stop}}

As discussed in Section \ref{sec:objective}, the main purpose of
the upper target $U_{{\scriptscriptstyle \!W}}$ in the performance
measure is to reduce downside risk. However, in multiperiod optimization,
a paradox might occur when the realized wealth overshoots the upper
target: by default, the portfolio optimizer might tell the fund manager
to pick the assets most likely to fall. It is trivial to see that,
when $W_{t}\geq U_{{\scriptscriptstyle \!W}}R_{f}^{-(T-t)}$, one
can outperform the upper target for certain by henceforth investing
$U_{{\scriptscriptstyle \!W}}R_{f}^{-(T-t)}$ amount of wealth into
the risk-free asset and taking out the balance amount $W_{t}-U_{{\scriptscriptstyle \!W}}R_{f}^{-(T-t)}$
from the problem. To implement such a correction, two approaches are
possible:
\begin{enumerate}
\item One can replace $T$ by $\min\{T,\tau\}$ in the value function in
equation \eqref{eq:objective}, where $\tau$ is the first (stopping)
time such that $W_{\tau}\geq U_{{\scriptscriptstyle \!W}}R_{f}^{-(T-\tau)}$.
At time $\tau$ (if it occurs before $T$), the dynamic optimization
stops: the amount $U_{{\scriptscriptstyle \!W}}R_{f}^{-(T-\tau)}$
is invested in the risk-free asset, and the balance amount $W_{\tau}-U_{{\scriptscriptstyle \!W}}R_{f}^{-(T-\tau)}$
is taken out. 
\item Alternatively, one can add an extra dynamic control to the problem:
dynamic withdrawal/consumption, see \citet*{Dang2017} for example. 
\end{enumerate}
For simplicity, we use the first approach in this paper. Based on
our numerical experiments, we find that imposing this stop-profit
rule does not significantly affect the terminal wealth distribution,
as usually only a very small portion of wealth realizations overshoot
the upper bound. For example, we show in the numerical section that
about 1\% of the realizations overshoot the upper bound for $[L_{\!{\scriptscriptstyle W}}=1.0,U_{{\scriptscriptstyle \!W}}=1.1]$,
and virtually 0\% for $[L_{\!{\scriptscriptstyle W}}=1.0,U_{{\scriptscriptstyle \!W}}=1.2]$.

\section{Extensions\label{sec:Extensions}}

This section adapts the two-stage LSMC method to alternative investment
objectives. We first describe how to use the two-stage LSMC method
to deal with the CRRA utility approach, then we adapt the formulation
of the STRS to the Flat Target Range Strategy (FTRS) which purely
maximizes the probability of achieving a prespecified target range
without further attempts to rally for profits, and to target range
strategies based on a stochastic benchmark, for which the absolute
fixed target range is replaced by a relative target range.

\subsection{CRRA utility\label{subsec:CRRA-utility}}

In the classical LSMC approach, a conditional expected utility of
the type $\mathbb{E}[\mathcal{U}(W_{T})|\mathbf{Z}_{t_{n}}=z,W_{t_{n}}=w]$
would be approximated by $\beta\cdot\psi(z,w)$, which may lead to
large numerical errors when the utility function $\mathcal{U}$ is
highly non-linear, see \citet{vanBinsbergen2007}, \citet{Garlappi2009},
\citet{Denault2017}, \citet{Zhang2018} and \citet{Andreasson2018}.
The proposed two-stage regression avoids this non-linearity problem
and greatly improves the stability of the LSMC method. In this subsection,
we derive the two-stage continuation value estimates for the CRRA
utility approach. These estimates involve the following special functions: 
\begin{itemize}
\item Gamma function: 
\[
\Gamma\left(z\right)=\int_{0}^{\infty}t^{z-1}\exp\left(-t\right)\text{d}t
\]
\item Rising factorial: 
\[
z^{(n)}=\frac{\Gamma\left(z+n\right)}{\Gamma\left(z\right)}
\]
\item Confluent hypergeometric function of the first kind: 
\begin{eqnarray*}
_{1}F_{1}\left(a,b,z\right) & = & \sum_{n=0}^{\infty}\frac{a^{(n)}}{b^{(n)}}\frac{z^{n}}{n!}
\end{eqnarray*}
\item Confluent hypergeometric function of the second kind: 
\begin{eqnarray*}
\Psi\left(a,b,z\right) & = & \frac{\Gamma\left(1-b\right)}{\Gamma\left(a-b+1\right)}{}_{1}F_{1}\left(a,b,z\right)+\frac{\Gamma\left(b-1\right)}{\Gamma\left(a\right)}z^{1-b}{}_{1}F_{1}\left(a-b+1,2-b,z\right)
\end{eqnarray*}
\end{itemize}
Assume that the conditional mean of the terminal wealth $\hat{\mu}_{t_{n}}^{j}(z,w)$
and the standard deviation $\hat{\sigma}_{t_{n}}^{j}$ have been estimated
according to equations \eqref{eq:ls} and \eqref{eq:linear-wealth}.
Then, using the general formula for the real moments of a Gaussian
distribution (\citet{Winkelbauer2014}), the continuation value function
in the CRRA utility approach is given by 
\begin{eqnarray}
\hat{\text{CV}}_{t_{n}}^{j}\left(z,w\right) & = & \mathbb{E}\left[\left.\frac{\left.\hat{W}_{t_{N}}\right.^{1-\gamma}}{1-\gamma}\right|\mathbf{Z}_{t_{n}}=z,W_{t_{n}}=w,\boldsymbol{\alpha}_{t_{n}}=\mathbf{a}_{j}\right]\nonumber \\
 & = & \frac{\left(\hat{\sigma}_{t_{n}}^{j}\right)^{1-\gamma}}{1-\gamma}\cdot\left(-i\sqrt{2}\right)^{1-\gamma}\cdot\Psi\left(-\frac{1-\gamma}{2},\frac{1}{2},-\frac{1}{2}\left(\frac{\hat{\mu}_{t_{n}}^{j}\left(z,w\right)}{\hat{\sigma}_{t_{n}}^{j}}\right)^{2}\right).\label{eq:cv-CRRA}
\end{eqnarray}

We use this closed-form formula for the numerical comparisons in Subsection
\ref{subsec:Model-validation}.

\subsection{Flat target range strategy \label{subsec:prob}}

\begin{figure}[H]
\caption{Flat Target Range Function\label{fig:EnC}}

\smallskip
\centering{}%
\begin{minipage}[t]{0.45\columnwidth}%
\includegraphics[scale=0.55]{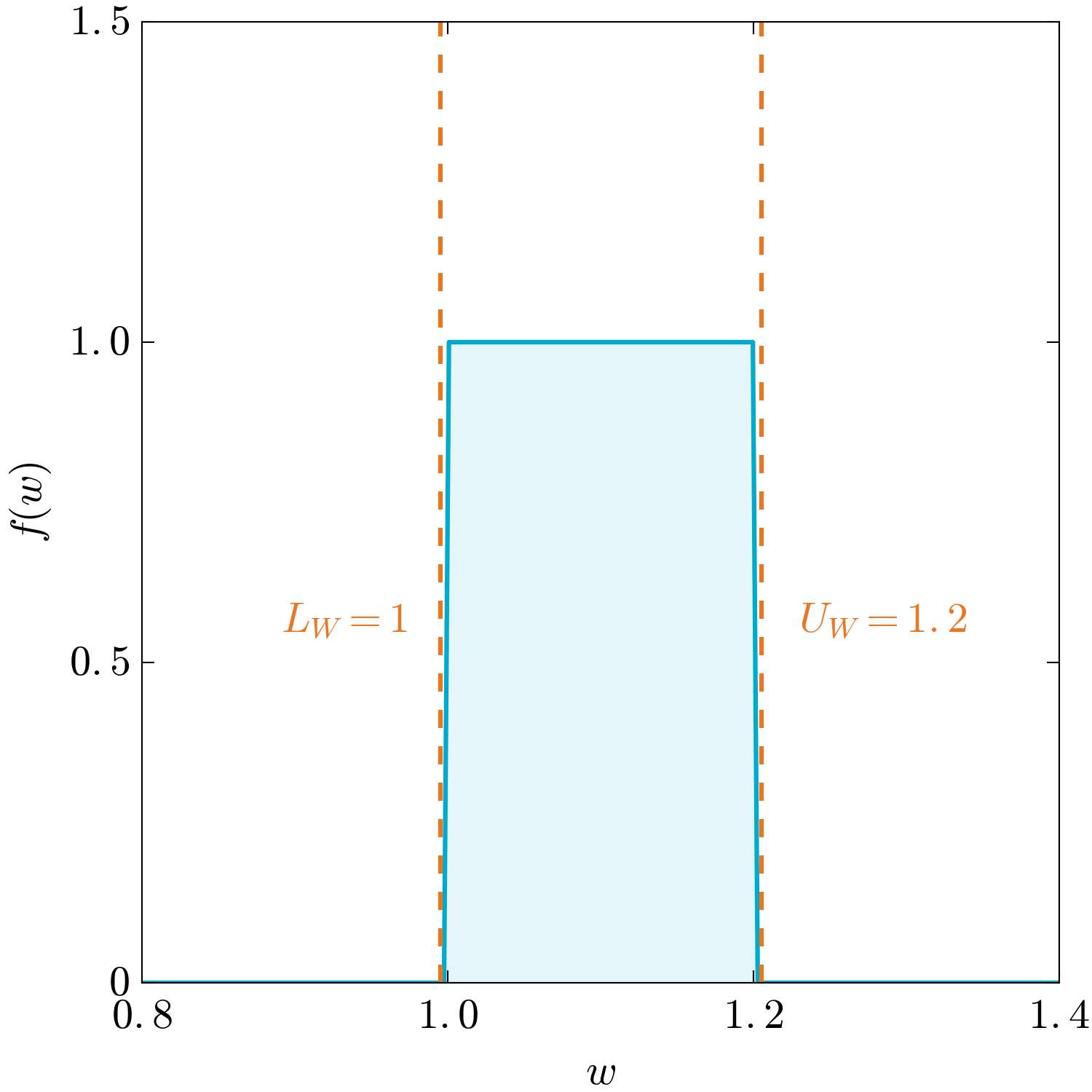}%
\end{minipage}
\end{figure}

The return distribution produced by the STRS \eqref{eq:f} is skewed
towards the upper return target. Yet, there exists some other types
of portfolio optimization problems (such as life-cycle and insurance-related
investments) for which the ability to remain solvent prevails over
the appetite for high expected return. For such problems, one can
adjust the skewed target range shape \eqref{eq:f} to a flat target
range shape given by 
\begin{equation}
f(w)=\mathbbm{1}\left\{ L_{{\scriptscriptstyle \!W}}\leq w\leq U_{{\scriptscriptstyle \!W}}\right\} .\label{eq:f_uniform}
\end{equation}
Figure \ref{fig:EnC} illustrates the above equation \eqref{eq:f_uniform}
with $[L_{{\scriptscriptstyle \!W}},U_{{\scriptscriptstyle \!W}}]=[1.0,1.2]$.

Then the portfolio optimization problem becomes 
\begin{eqnarray}
v_{t}(z,w) & = & \sup_{\left\{ \boldsymbol{\alpha}_{\tau}\in\mathcal{A}\right\} _{t\leq\tau\leq T}}\mbox{\ensuremath{\mathbb{E}}}\left[\mathbbm{1}\left\{ L_{{\scriptscriptstyle \!W}}\leq w\leq U_{{\scriptscriptstyle \!W}}\right\} \left|\mathbf{Z}_{t}=z,W_{t}=w\right.\right]\nonumber \\
 & = & \sup_{\left\{ \boldsymbol{\alpha}_{\tau}\in\mathcal{A}\right\} _{t\leq\tau\leq T}}\mathbb{P}\left[L_{{\scriptscriptstyle \!W}}\leq W_{T}\leq U_{{\scriptscriptstyle \!W}}\left|\mathbf{Z}_{t}=z,W_{t}=w\right.\right],\label{eq:prob-max}
\end{eqnarray}
which is a pure probability maximizing strategy.

The conservative FTRS can be deemed more flexible than the classical
Value-at-Risk (VaR) minimization approach: when $U_{{\scriptscriptstyle \!W}}=+\infty$,
the FTRS \eqref{eq:prob-max} and VaR minimization achieve comparable
investment outcomes, the difference being a fixed, absolute cut-off
level for the former and an implicit, relative cut-off level for the
latter. In particular, the FTRS minimizes the probability of being
below a particular loss level, while the VaR procedure minimizes a
particular loss quantile. When $U_{{\scriptscriptstyle \!W}}$ is
finite, the FTRS provides greater flexibility for investors to devise
their risk preferences, as the lower return target $L_{{\scriptscriptstyle \!W}}$
in such circumstances is an explicit input from the investor, and
the option to fix an upper target $U_{{\scriptscriptstyle \!W}}$
broadens the range of possible risk profiles.

Assuming that the conditional mean of the terminal wealth $\hat{\mu}_{t_{n}}^{j}(z,w)$
and the standard deviation $\hat{\sigma}_{t_{n}}^{j}$ have been estimated
according to equations \eqref{eq:ls} and \eqref{eq:linear-wealth},
the continuation value function is simply given by 
\begin{eqnarray}
\hat{\text{CV}}{}_{t_{n}}^{j}\left(z,w\right) & = & \mathbb{P}\left[\left.\mathbbm{1}\left\{ L_{{\scriptscriptstyle \!W}}\leq W_{t_{N}}\leq U_{{\scriptscriptstyle \!W}}\right\} \right|\mathbf{Z}_{t_{n}}=z,W_{t_{n}}=w,\boldsymbol{\alpha}_{t_{n}}=\mathbf{a}_{j}\right]\nonumber \\
 & = & \mathbb{P}_{\varepsilon}\left[\mathbbm{1}\left\{ L_{{\scriptscriptstyle \!W}}\leq\hat{\mu}_{t_{n}}^{j}\left(z,w\right)+\hat{\sigma}_{t_{n}}^{j}\varepsilon\leq U_{{\scriptscriptstyle \!W}}\right\} \right]\nonumber \\
 & = & \Phi\left(\frac{U_{{\scriptscriptstyle \!W}}-\hat{\mu}_{t_{n}}^{j}\left(z,w\right)}{\hat{\sigma}_{t_{n}}^{j}}\right)-\Phi\left(\frac{L_{{\scriptscriptstyle \!W}}-\hat{\mu}_{t_{n}}^{j}\left(z,w\right)}{\hat{\sigma}_{t_{n}}^{j}}\right).
\end{eqnarray}

\subsection{Target range over a stochastic benchmark\label{subsec:relative}}

It is also possible to define the return thresholds $L_{{\scriptscriptstyle \!W}}$
and $U_{{\scriptscriptstyle \!W}}$ relatively to a stochastic benchmark,
be it stock market index, inflation rate, exchange rate or interest
rate. We refer to \citet{Franks1992}, \citet{Browne1999a}, \citet{Brogan2005}
and \citet{Gaivoronski2005} for classical investment strategies that
aim to outperform a stochastic benchmark.

Denote by $B$ the stochastic benchmark of interest, and define the
relative excess wealth as $W-B$. We can then modify the target range
function as: 
\begin{equation}
f_{B}(w,b):=(w-b)\mathbbm{1}\{L_{{\scriptscriptstyle \!W}}\leq w-b\leq U_{{\scriptscriptstyle \!W}}\}\,,\label{eq:f_benchmark}
\end{equation}
for STRS, and 
\begin{equation}
f_{B}(w,b):=\mathbbm{1}\{L_{{\scriptscriptstyle \!W}}\leq w-b\leq U_{{\scriptscriptstyle \!W}}\}\,,\label{eq:f_flat_benchmark}
\end{equation}
for FTRS. 

The stochastic benchmark $B$ can be simply modeled as one additional
exogenous state variable, so that this new problem can be solved using
the same approach developed in Section \ref{sec:LSMC}.

\section{Numerical Experiments\label{sec:Numerical}}

In this section, we test the skewed target range strategy (STRS),
and illustrate how it can achieve the investor's range objective.
Table \ref{tab:asset-class} summarizes the asset classes and the
exogenous state variables used for our numerical experiments. We consider
a portfolio invested in five assets: risk-free cash, U.S. bonds (AGG),
U.S. shares (SPY), international shares (IFA) and emerging market
shares (EEM), the other assets listed in Table \ref{tab:asset-class}
being used as return predictors. 

\begin{table}[H]
\caption{\label{tab:asset-class}Risky assets and return predictors}

\smallskip
\begin{singlespace}
\centering{}%
\begin{tabular}{lllll}
{\footnotesize{}Assets } &  & {\footnotesize{}Underlying} &  & {\footnotesize{}Data source}\tabularnewline
\hline 
{\footnotesize{}U.S. Bonds } &  & {\footnotesize{}AGG (ETF) } &  & {\footnotesize{}Yahoo Finance}\tabularnewline
{\footnotesize{}U.S. Shares } &  & {\footnotesize{}SPY (ETF) } &  & {\footnotesize{}Yahoo Finance}\tabularnewline
{\footnotesize{}International Shares } &  & {\footnotesize{}IFA (ETF) } &  & {\footnotesize{}Yahoo Finance}\tabularnewline
{\footnotesize{}Emerging Market Shares } &  & {\footnotesize{}EEM (ETF) } &  & {\footnotesize{}Yahoo Finance}\tabularnewline
{\footnotesize{}Japanese shares} &  & {\footnotesize{}NIKKEI225} &  & {\footnotesize{}Yahoo Finance}\tabularnewline
{\footnotesize{}U.K. shares} &  & {\footnotesize{}FTSE100} &  & {\footnotesize{}Yahoo Finance}\tabularnewline
{\footnotesize{}Australian shares} &  & {\footnotesize{}ASX200} &  & {\footnotesize{}Yahoo Finance}\tabularnewline
{\footnotesize{}Gold } &  & {\footnotesize{}Spot Price} &  & {\footnotesize{}World Gold Council}\tabularnewline
{\footnotesize{}Crude Oil } &  & {\footnotesize{}Spot Price} &  & {\footnotesize{}U.S. Energy Info. Admin.}\tabularnewline
{\footnotesize{}U.S. Dollar } &  & {\footnotesize{}USD Index } &  & {\footnotesize{}Federal Reserve}\tabularnewline
{\footnotesize{}Japanese Yen} &  & {\footnotesize{}JPYUSD} &  & {\footnotesize{}Federal Reserve}\tabularnewline
{\footnotesize{}Euro} &  & {\footnotesize{}USDEUR} &  & {\footnotesize{}Federal Reserve}\tabularnewline
{\footnotesize{}Australian Dollar} &  & {\footnotesize{}USDAUD} &  & {\footnotesize{}Federal Reserve}\tabularnewline
\hline 
\end{tabular}
\end{singlespace}
\end{table}

The annual interest rate on the cash component is set to be $2\%$.
We assume $0.1\%$ proportional transaction costs and we refer to
\citet{Zhang2018} on how to deal with switching costs in the LSMC
algorithm with endogenous variables. A first-order vector autoregression
model is calibrated to the monthly log-returns of the assets listed
in Table \ref{tab:asset-class} from September 2003 to March 2016.
By bootstrapping the residuals, 10,000 simulation paths are generated
for one year with monthly time steps. The two-stage regression method
approximates a linear wealth $W_{T}$, but not a concave utility $\mathcal{U}(W_{T})$;
as a result, a sample of 10,000 paths can be deemed sufficient to
reach numerical stability, as reported in \citet{vanBinsbergen2007}
and \citet{Zhang2018}. For the same reason, we use a simple second-order
multivariate polynomial as the basis functions for the linear least
squares regressions in the algorithm. For simplicity, all the reported
distributions are simulated in-sample, which might in theory make
the estimation upward-biased. In the numerical experiments, we use
a mesh of 0.2 increment for the discrete control grid and we do not
allow short-selling and borrowing. Apart from Subsection \ref{subsec:Model-validation}
where a state-dependent standard deviation is tested, the state-independent
standard deviation is used for all the other numerical experiments.
The program is coded in Python 3.4.3, and it takes approximately two
hours on a 2.2 GHz Intel Core i7 CPU to complete the computation for
$M=10,000$ paths, 12 time steps, 13 state variables, a second-order
polynomial basis, and a control mesh of 0.2 for a five-dimensional
portfolio. 

\subsection{Wealth distribution}

\begin{figure}[H]
\caption{Terminal wealth distribution using STRS\label{fig:target}}

\medskip
\begin{centering}
\begin{minipage}[t]{0.4\columnwidth}%
\includegraphics[scale=0.55]{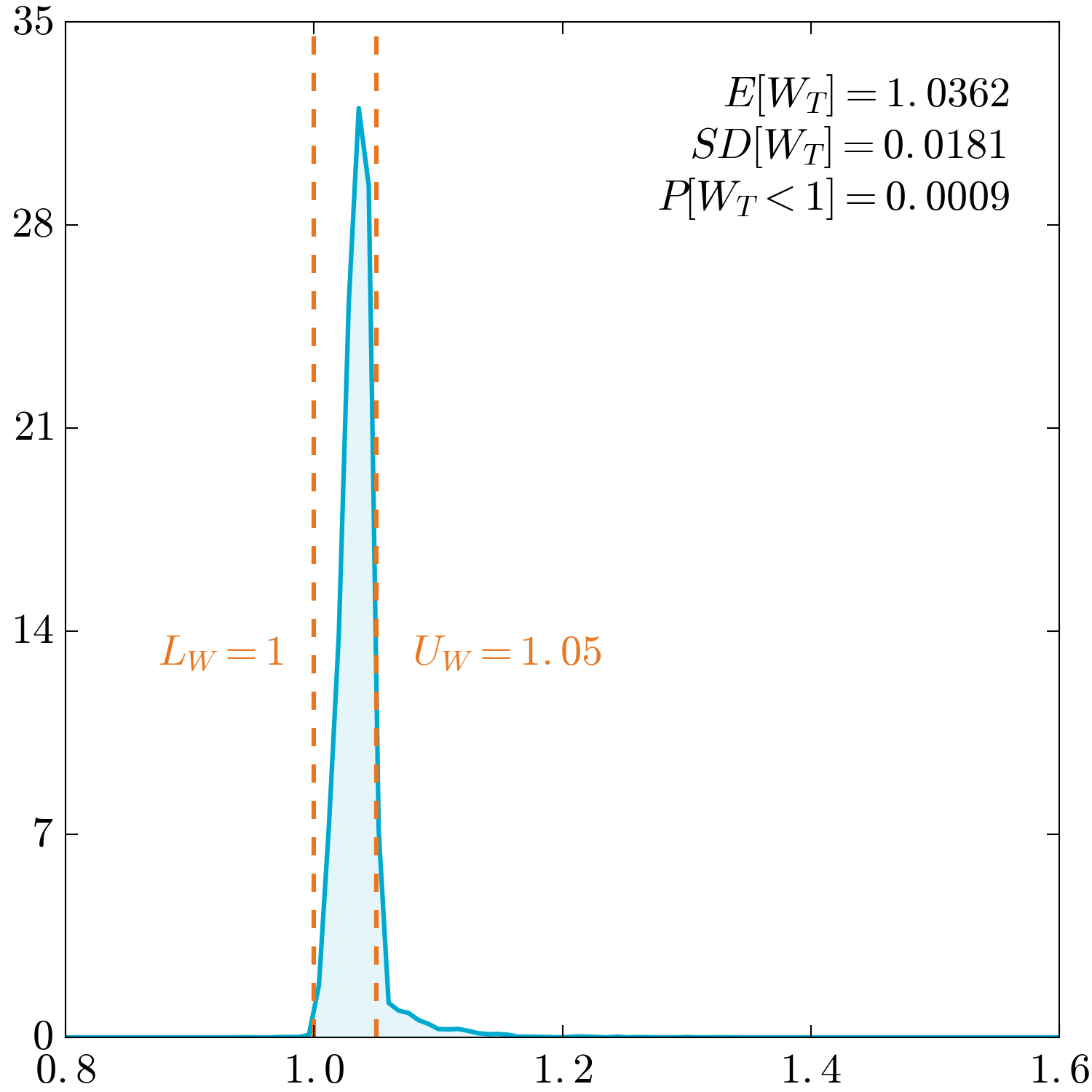}%
\end{minipage}\qquad{}%
\begin{minipage}[t]{0.4\columnwidth}%
\includegraphics[scale=0.55]{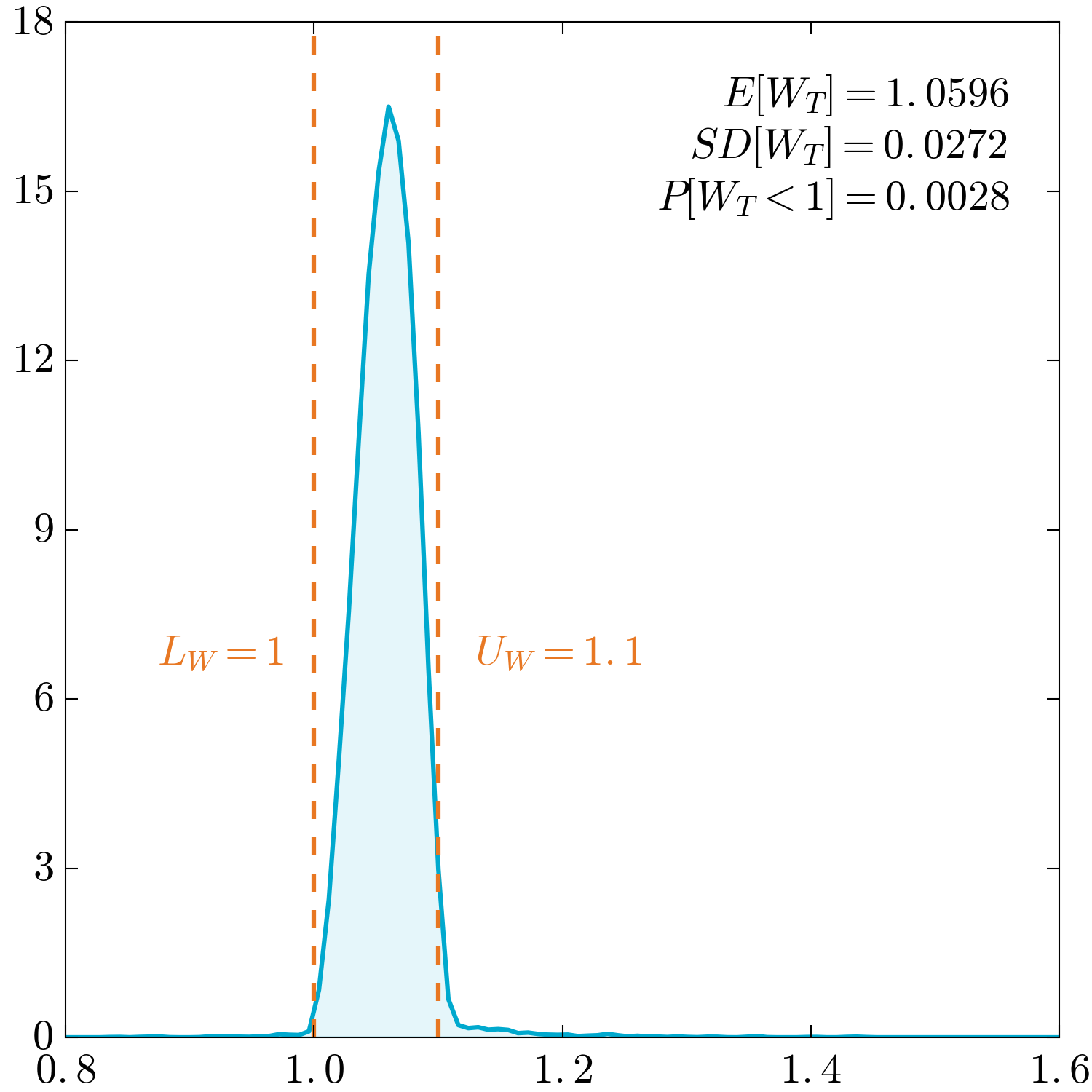}%
\end{minipage}
\par\end{centering}
\centering{}%
\begin{minipage}[t]{0.4\columnwidth}%
\includegraphics[scale=0.55]{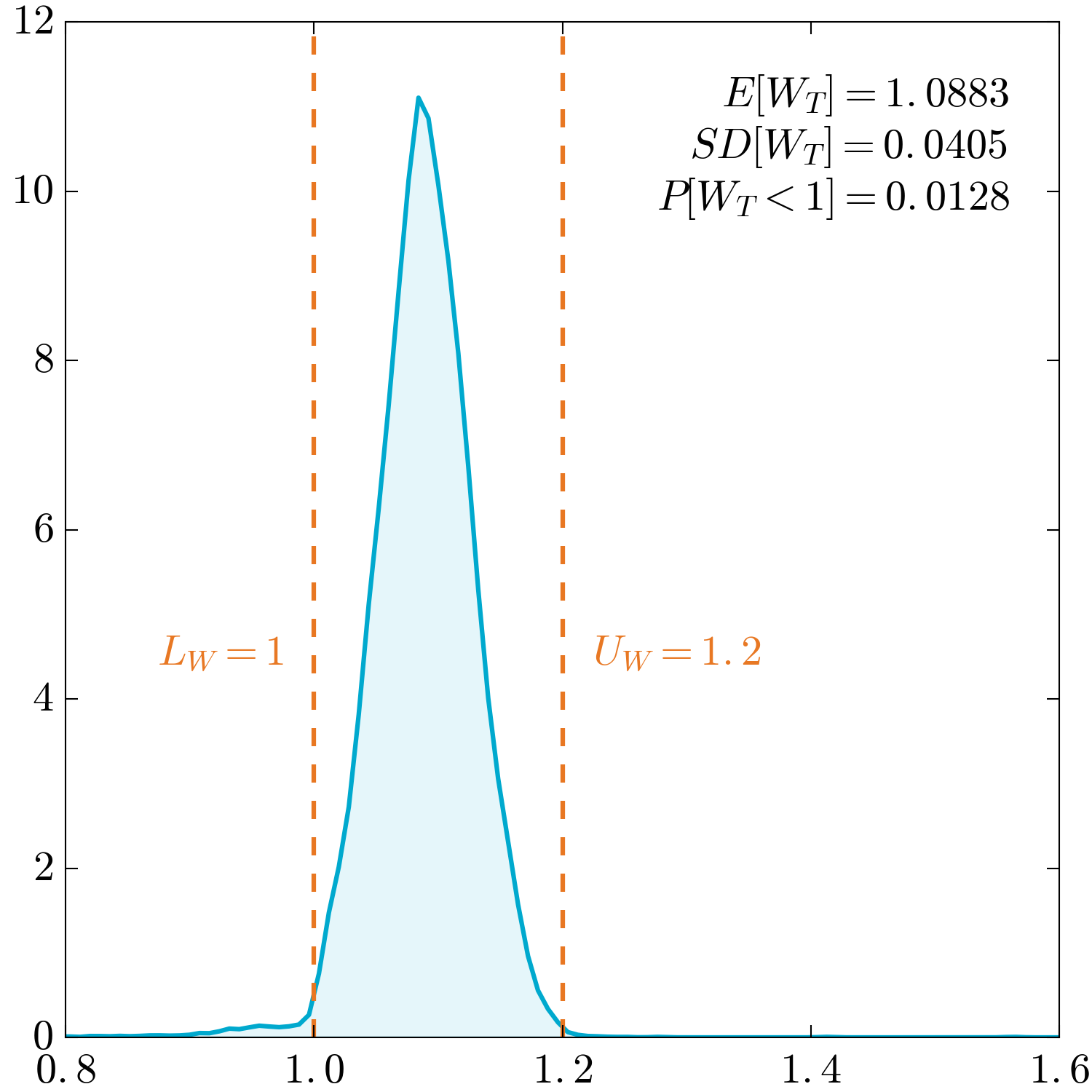}%
\end{minipage}\qquad{}%
\begin{minipage}[t]{0.4\columnwidth}%
\includegraphics[scale=0.55]{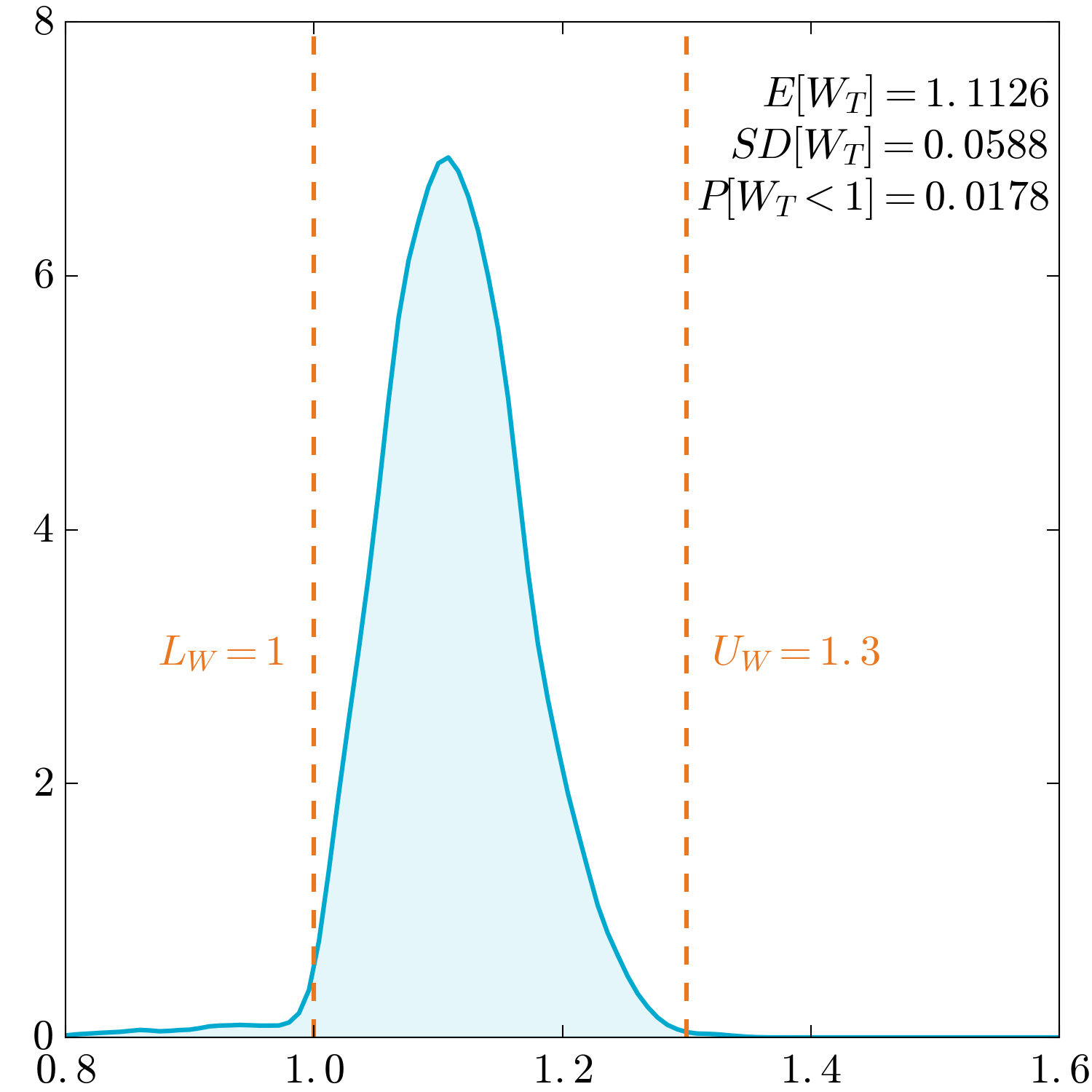}%
\end{minipage}
\end{figure}

Figure \ref{fig:target} provides some examples of estimated distribution
of terminal portfolio value when using the STRS. We recall that the
portfolio value $W$ and the bounds $[L_{{\scriptscriptstyle \!W}},U_{{\scriptscriptstyle \!W}}]$
are scaled by the initial wealth, so that without loss of generality
we assume $W_{0}=1.00$. The lower target $L_{{\scriptscriptstyle \!W}}$
is set to the initial wealth level $1.00$, a natural choice representing
the preference of investors for capital protection. Four different
upper targets $U_{{\scriptscriptstyle \!W}}$ are tested: $1.05$,
$1.10$, $1.20$ and $1.30$. 

Several comments can be made about the shape of the terminal wealth
distribution produced by the STRS in Figure \ref{fig:target}. The
most striking observation is that the STRS does confine most of the
wealth realizations within the predefined target range, and for low
upper target levels $U_{{\scriptscriptstyle \!W}}=1.05$ and $U_{{\scriptscriptstyle \!W}}=1.10$,
the wealth distributions mimics to some extent the shape of the skewed
target range function \eqref{eq:f}, making downside risk negligible.
This suggests the two-stage LSMC algorithm is indeed capable of handling
an abrupt discontinuous payoff function properly. There are some wealth
realizations lying above the upper bound, which, in spite of the first
correction described in Subsection \ref{subsec:stop}, may occur due
to the discrete-time nature of monthly rebalancing (a large upward
jump can occur during one single month, after which the risky investment
is immediately stopped as described in Subsection \ref{subsec:stop}).
\begin{figure}[H]
\begin{centering}
\caption{Time evolution of wealth distribution using STRS\label{fig:evolution}}
\par\end{centering}
\smallskip
\begin{centering}
\hspace{-2em}%
\begin{minipage}[t]{0.4\textwidth}%
\includegraphics[scale=0.14]{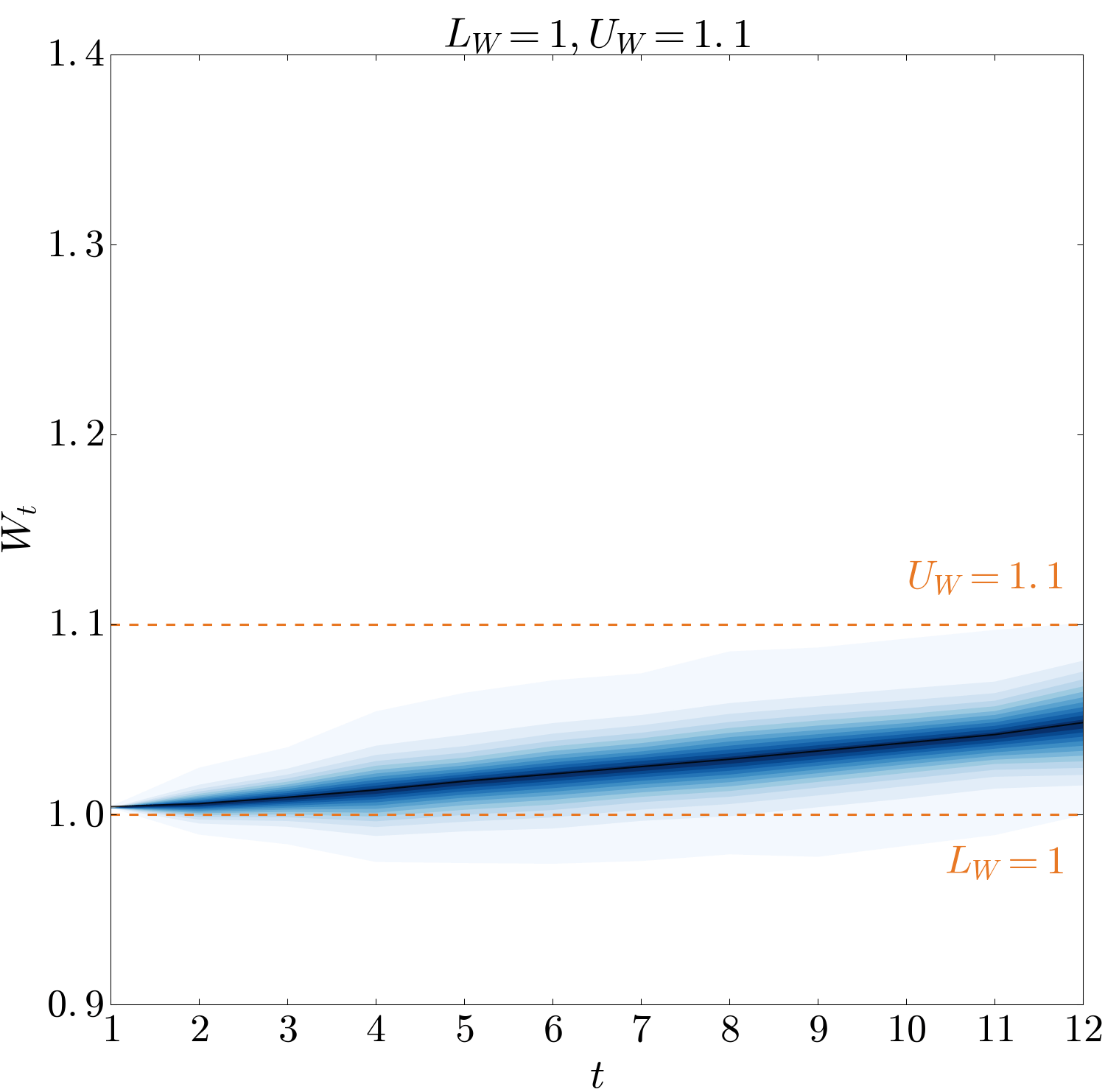}%
\end{minipage}\hspace{3em}%
\begin{minipage}[t]{0.4\textwidth}%
\includegraphics[scale=0.14]{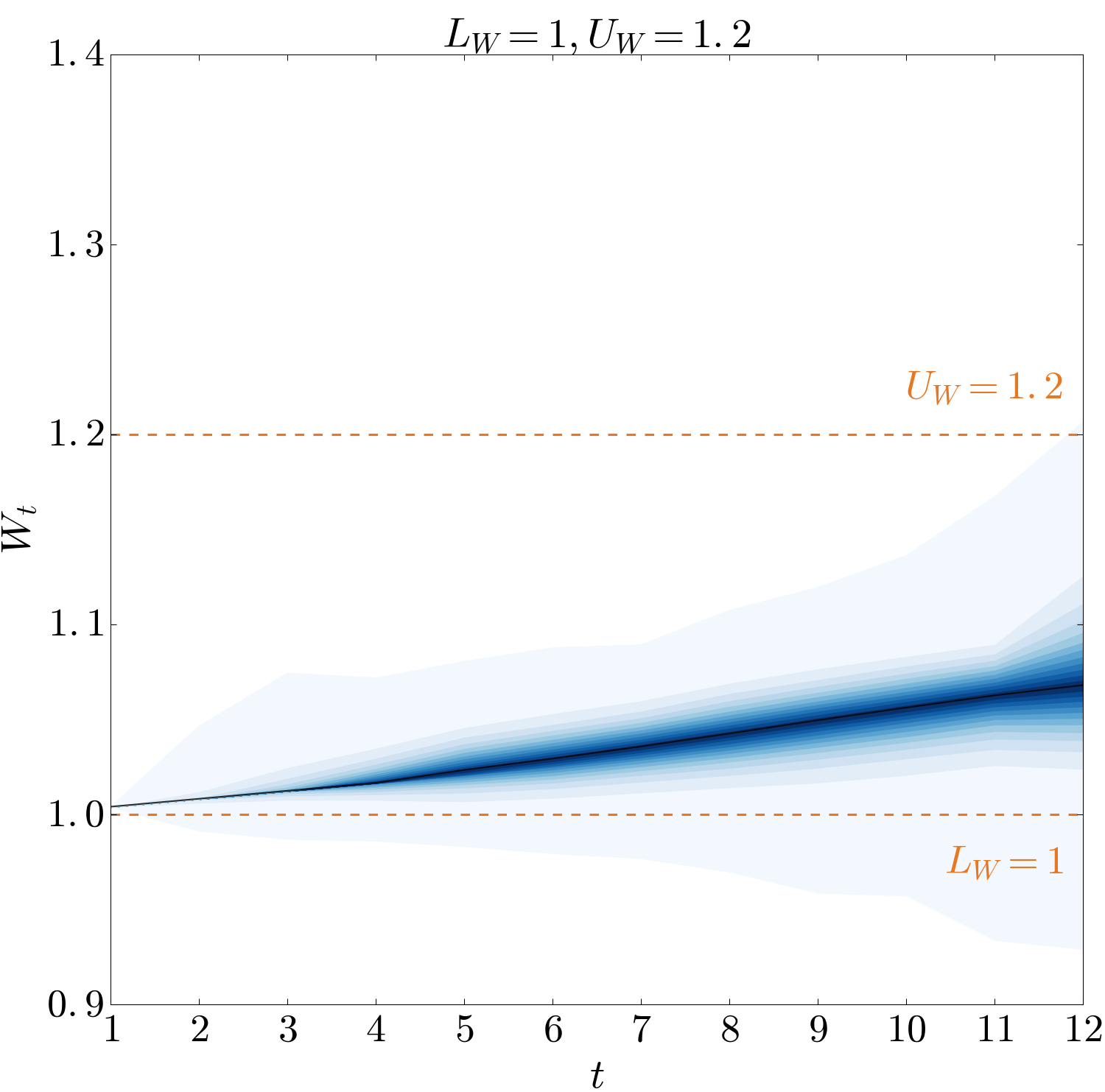}%
\end{minipage}
\par\end{centering}
\centering{}\hspace{-2em}%
\begin{minipage}[t]{0.4\columnwidth}%
\includegraphics[scale=0.14]{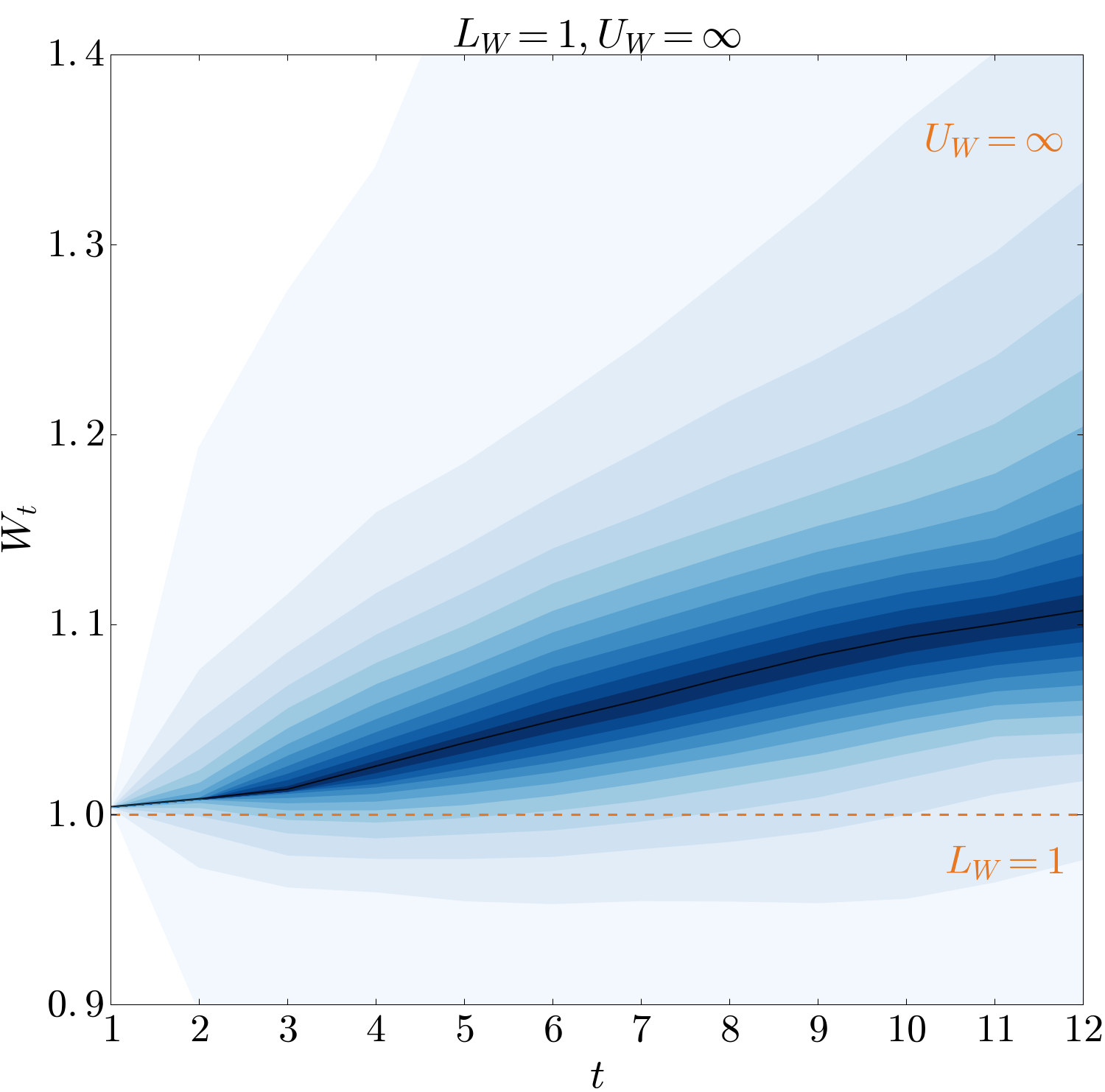}%
\end{minipage}\hspace{3em}%
\begin{minipage}[t]{0.4\columnwidth}%
\includegraphics[scale=0.14]{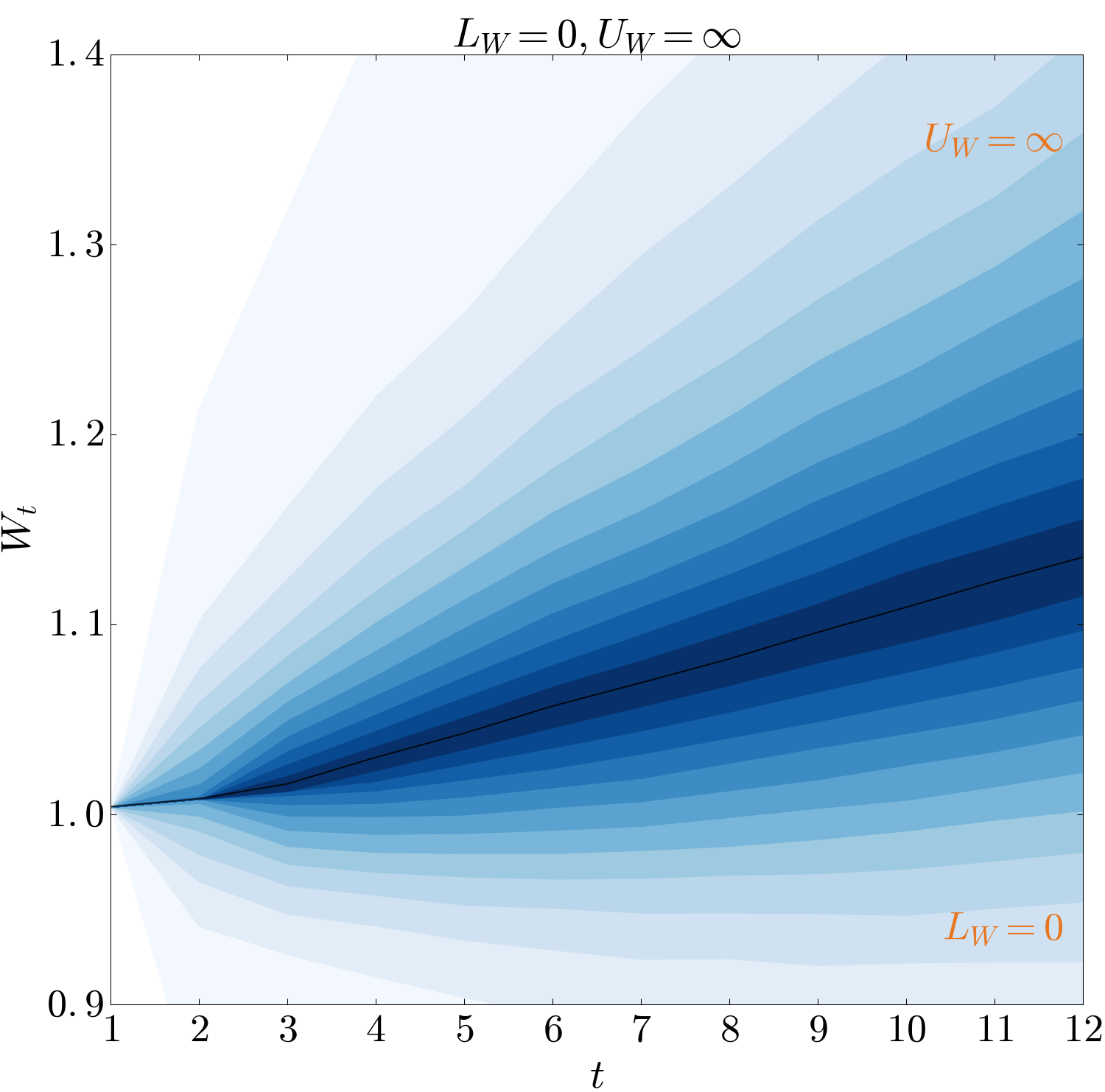}%
\end{minipage}
\end{figure}

As expected, setting the upper target $U_{{\scriptscriptstyle \!W}}$
to a higher level produces a higher expected terminal wealth with
higher standard deviation and greater downside risk (as measured by
the probability of losing capital). At the same time, the higher the
upper target $U_{{\scriptscriptstyle \!W}}$, the harder it is for
the terminal wealth distribution to be skewed towards the upper target.
Regarding the tails beyond the targeted range, the two low upper target
levels $U_{{\scriptscriptstyle \!W}}=1.05$ and $U_{{\scriptscriptstyle \!W}}=1.10$
produce larger right tails, while the two higher levels $U_{{\scriptscriptstyle \!W}}=1.20$
and $U_{{\scriptscriptstyle \!W}}=1.30$ produce larger left tails,
which is consistent with the fact that the greater $U_{{\scriptscriptstyle \!W}}$,
the higher the risk that the investor is willing to take to achieve
a higher return. This illustrates the capability of the STRS to cater
to different risk appetites.

An interesting quantity to monitor is the ratio $\mathcal{R}:=(\mathbb{E}[W_{T}]-L_{{\scriptscriptstyle \!W}})/(U_{{\scriptscriptstyle \!W}}-L_{{\scriptscriptstyle \!W}})$
which measures the location of the expected performance $\mathbb{E}[W_{T}]$
relative to the targeted range: $\mathcal{R}=0\%$ means $\mathbb{E}[W_{T}]=L_{{\scriptscriptstyle \!W}}$,
while at the opposite $\mathcal{R}=100\%$ means $\mathbb{E}[W_{T}]=U_{{\scriptscriptstyle \!W}}$.
In our experiments from Figure \ref{fig:target}, $\mathcal{R}$ is
a decreasing function of $U_{{\scriptscriptstyle \!W}}$, from $\mathcal{R}=72\%$
for $U_{{\scriptscriptstyle \!W}}=1.05$ down to $\mathcal{R}=38\%$
for $U_{{\scriptscriptstyle \!W}}=1.30$. This illustrates the natural
fact that the higher the desired upper target, the harder it is to
achieve it. One visible drawback of the proposed strategy is the relatively
long left tail when both the upper and lower targets are set to relatively
high levels, for example, $L_{{\scriptscriptstyle \!W}}\geq1.00$
and $U_{{\scriptscriptstyle \!W}}\geq1.20$. 

Figure \ref{fig:evolution} shows the time evolution of the wealth
distribution (0.05 percentile to 99.95 percentile) over the whole
investment horizon, for the STRS with $[L_{{\scriptscriptstyle \!W}}=1.0,U_{\negmedspace{\scriptscriptstyle W}}=1.1]$
(top-left panel), $[L_{{\scriptscriptstyle \!W}}=1.0,U_{\negmedspace{\scriptscriptstyle W}}=1.2]$
(top-right panel), $[L_{{\scriptscriptstyle \!W}}=1.0,U_{\negmedspace{\scriptscriptstyle W}}=\infty]$
(bottom-left panel) and $[L_{{\scriptscriptstyle \!W}}=0,U_{\negmedspace{\scriptscriptstyle W}}=\infty]$
(bottom-right panel), where the last strategy is equivalent to maximizing
the expected terminal wealth without taking risk into account. The
results show that the wealth distributions in the top panel are well
tightened within the prespecified target ranges over the whole investment
process, as opposed to the case $U_{\negmedspace{\scriptscriptstyle W}}=\infty$
in the bottom panel. Once again, as upside potential and downside
risk are naturally intertwined, one cannot protect against downside
risk very well when the upper target is set to a very high level,
as shown by the $[L_{{\scriptscriptstyle \!W}}=1.0,U_{\negmedspace{\scriptscriptstyle W}}=\infty]$
example (bottom-left panel).

\subsection{Sensitivity analysis and choice of $L_{{\scriptscriptstyle \!W}}$\label{subsec:Sensitivity}}

\begin{figure}[H]
\caption{Sensitivity analysis w.r.t. target bounds\label{fig:sensitivity}}

\vspace{1em}

\begin{centering}
\begin{minipage}[t]{0.4\columnwidth}%
\includegraphics[scale=0.13]{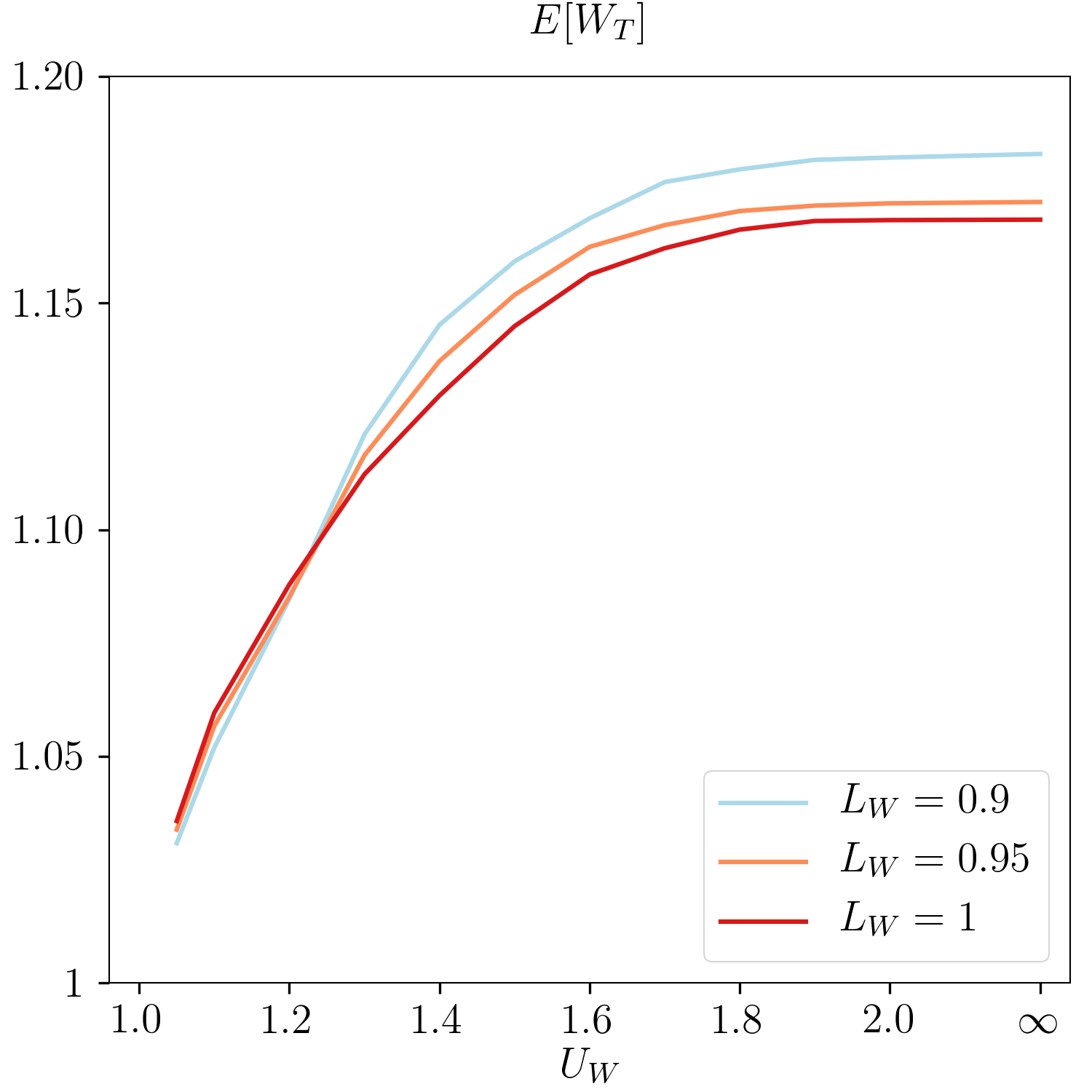}%
\end{minipage}\qquad{}%
\begin{minipage}[t]{0.4\columnwidth}%
\includegraphics[scale=0.13]{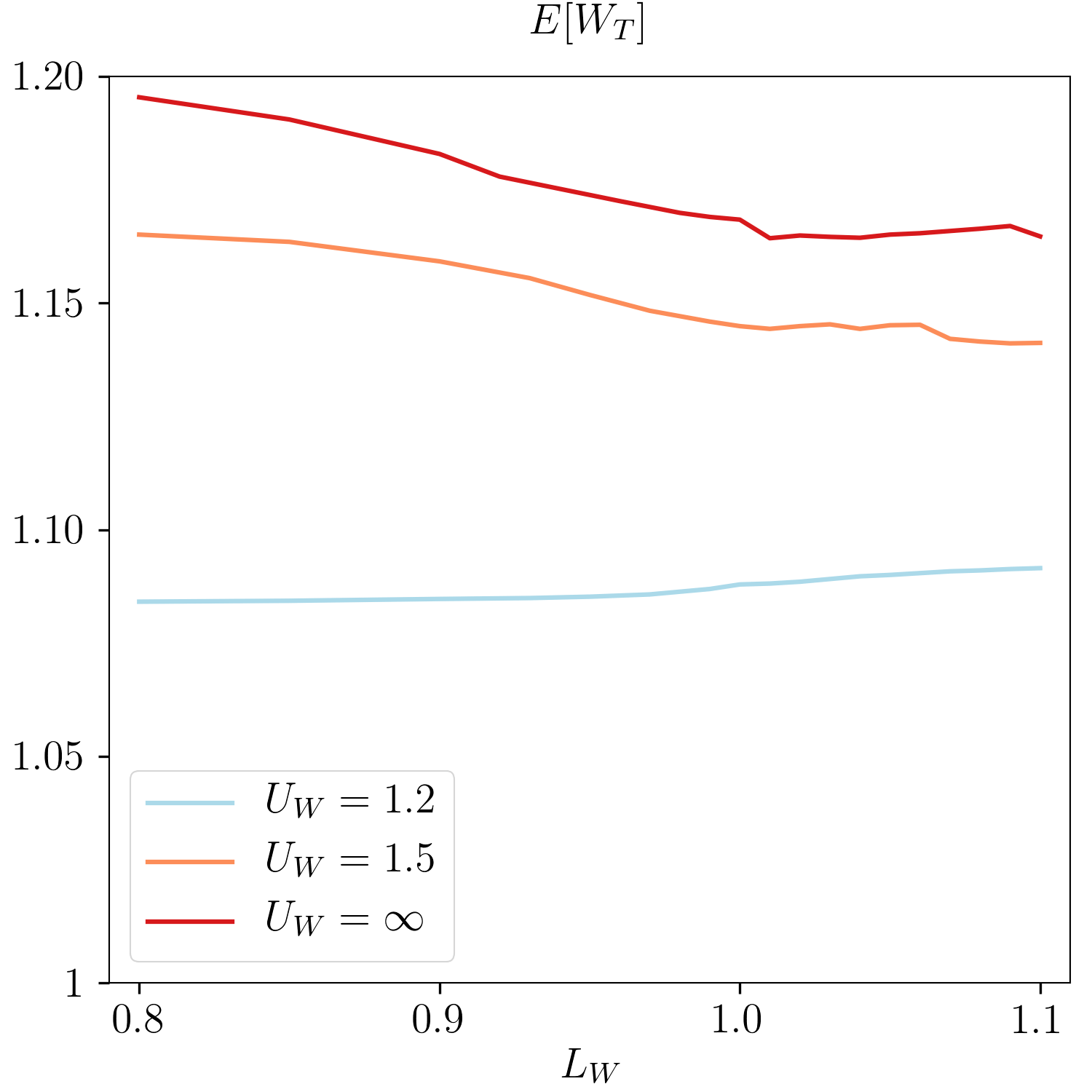}%
\end{minipage}
\par\end{centering}
\vspace{1em}

\begin{centering}
\begin{minipage}[t]{0.4\columnwidth}%
\includegraphics[scale=0.13]{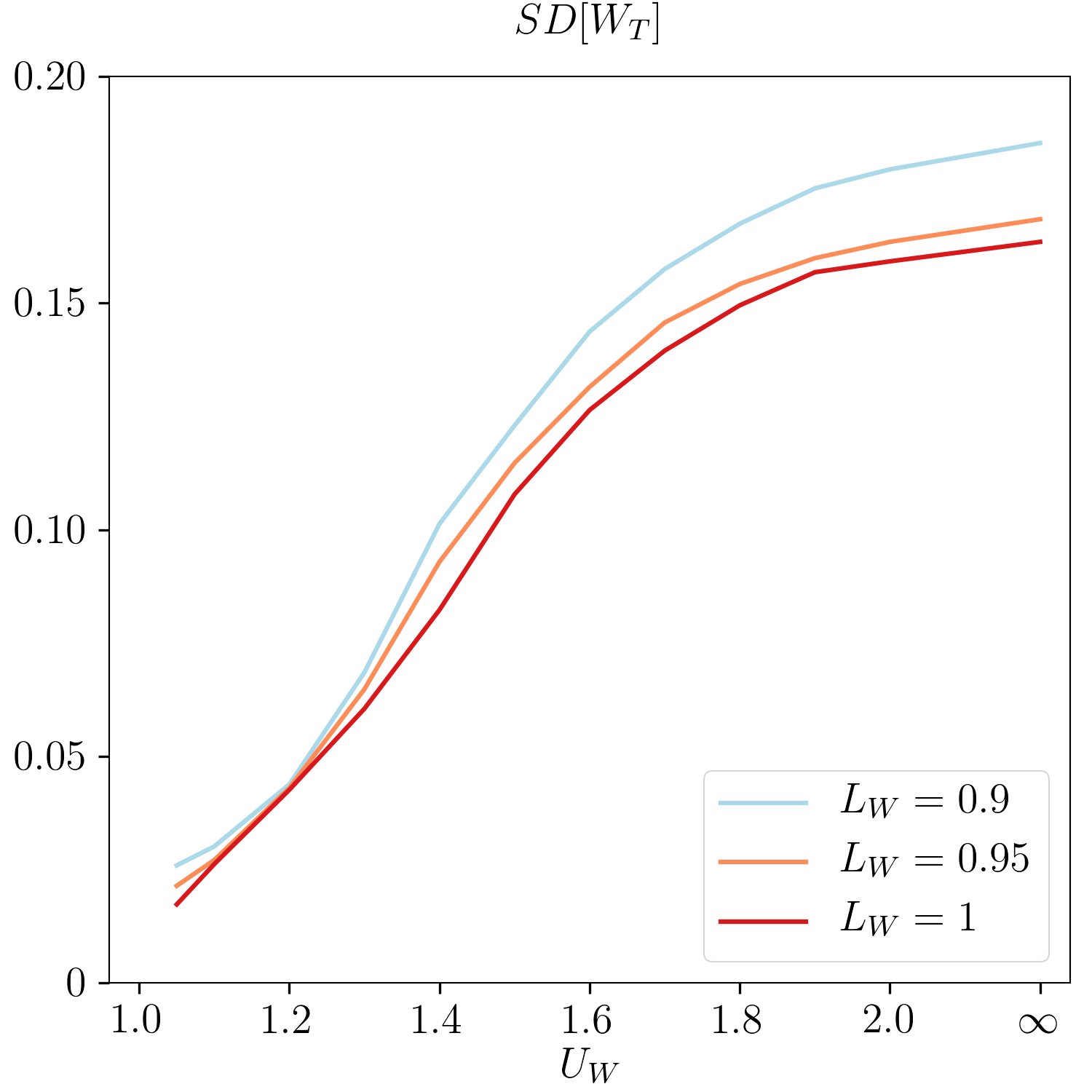}%
\end{minipage}\qquad{}%
\begin{minipage}[t]{0.4\columnwidth}%
\includegraphics[scale=0.13]{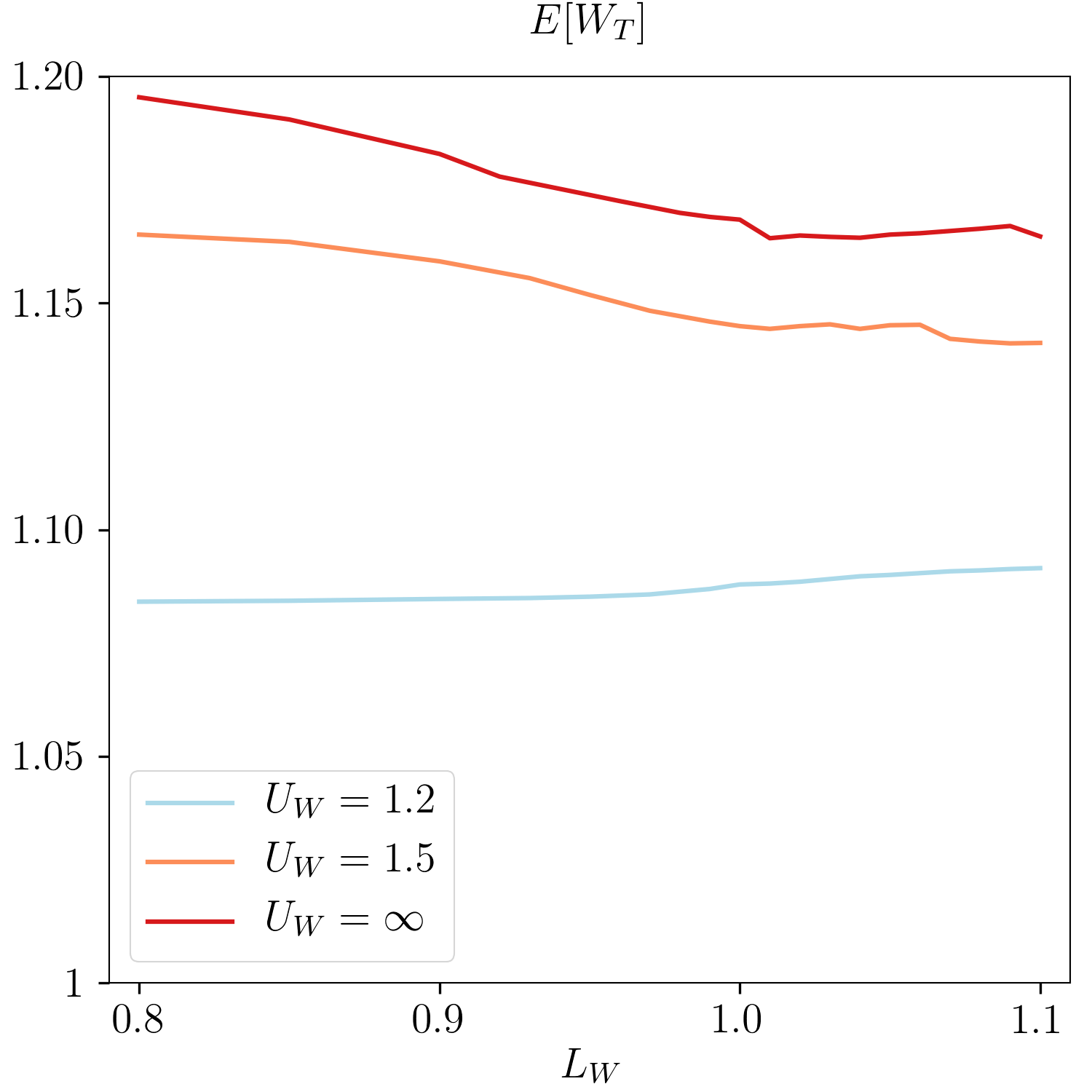}%
\end{minipage}
\par\end{centering}
\vspace{1em}

\centering{}%
\begin{minipage}[t]{0.4\columnwidth}%
\includegraphics[scale=0.13]{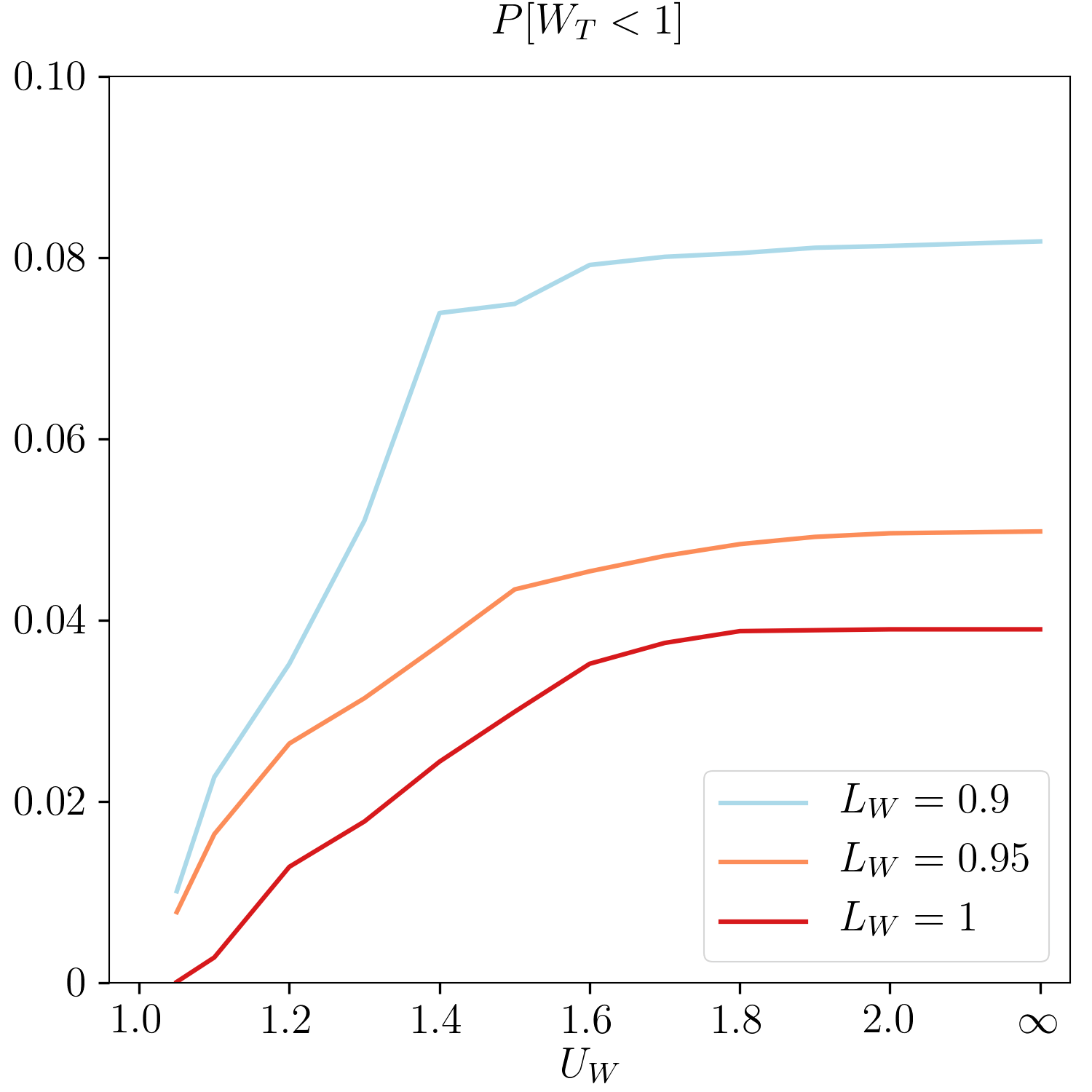}%
\end{minipage}\qquad{}%
\begin{minipage}[t]{0.4\columnwidth}%
\includegraphics[scale=0.13]{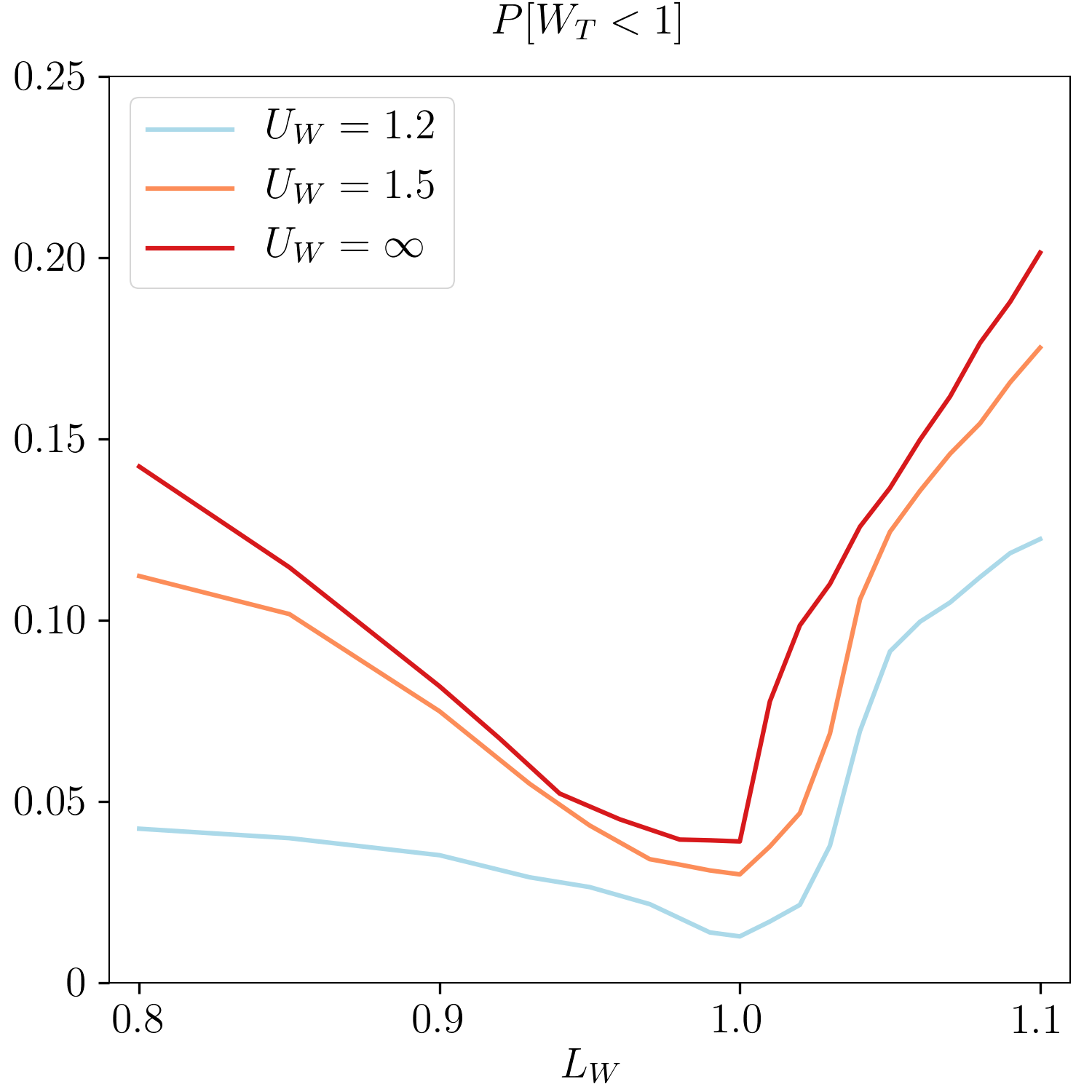}%
\end{minipage}
\end{figure}

The next experiment is a sensitivity analysis of the expected terminal
wealth, standard deviation and downside risk with respect to the bounds
of the STRS. Figure \ref{fig:sensitivity} shows how the expected
terminal wealth ($\mathbb{E}[W_{T}]$, first row), the standard deviation
of the terminal wealth ($\text{SD}[W_{T}]$, second row) and the downside
risk ($\mathbb{P}[W_{T}<1]$, third row) are affected by changes in
the upper bound $U_{\!{\scriptscriptstyle W}}$ (left column) and
by changes in the lower bound $L_{\!{\scriptscriptstyle W}}$ (right
column).

The left column of Figure \ref{fig:sensitivity} shows how the expectation
$\mathbb{E}[W_{T}]$, standard deviation $\text{SD}[W_{T}]$ and downside
risk $\mathbb{P}[W_{T}<1]$ increase with $U_{{\scriptscriptstyle \!W}}$,
though a plateau is reached around $U_{\!{\scriptscriptstyle W}}=1.5$
for $\mathbb{P}[W_{T}<1]$ and around $U_{\!{\scriptscriptstyle W}}=1.8$
for $\mathbb{E}[W_{T}]$. 

On the right column, one can see that the standard deviation $\text{SD}[W_{T}]$
and downside risk $\mathbb{P}[W_{T}<1]$ both increase when $L_{{\scriptscriptstyle \!W}}$
moves away from the initial wealth $W_{0}=1.0$. When $L_{{\scriptscriptstyle \!W}}>1.0$,
both risk measures increase with $|L_{{\scriptscriptstyle \!W}}-W_{0}|$
due to the additional risk required at the beginning of the trading
period to force the portfolio value to grow from $W_{0}=1.0$ to the
lower target $L_{{\scriptscriptstyle \!W}}>W_{0}=1.0$. When $L_{{\scriptscriptstyle \!W}}<1.0$,
both risk measures also increase with $|W_{0}-L_{{\scriptscriptstyle \!W}}|$
due to the lack of immediate loss penalization. Nevertheless, the
net effect of $L_{{\scriptscriptstyle \!W}}$ on $\mathbb{E}[W_{T}]$
is mostly negligible. As a result, these observations suggest that
$L_{{\scriptscriptstyle \!W}}=W_{0}=1.0$ is an appropriate choice
for the lower bound of the targeted interval, from which the upper
bound $U_{\!{\scriptscriptstyle W}}$ can be set according to the
risk preference and the return requirement of the investor.

\subsection{Model validation\label{subsec:Model-validation}}

\begin{figure}[H]
\caption{Two-stage LSMC v.s. classical LSMC for CRRA utility\label{fig:LSMC-2LSMC-CRRA}}

\smallskip
\centering{}%
\begin{minipage}[t]{1\columnwidth}%
\begin{center}
\includegraphics[scale=0.13]{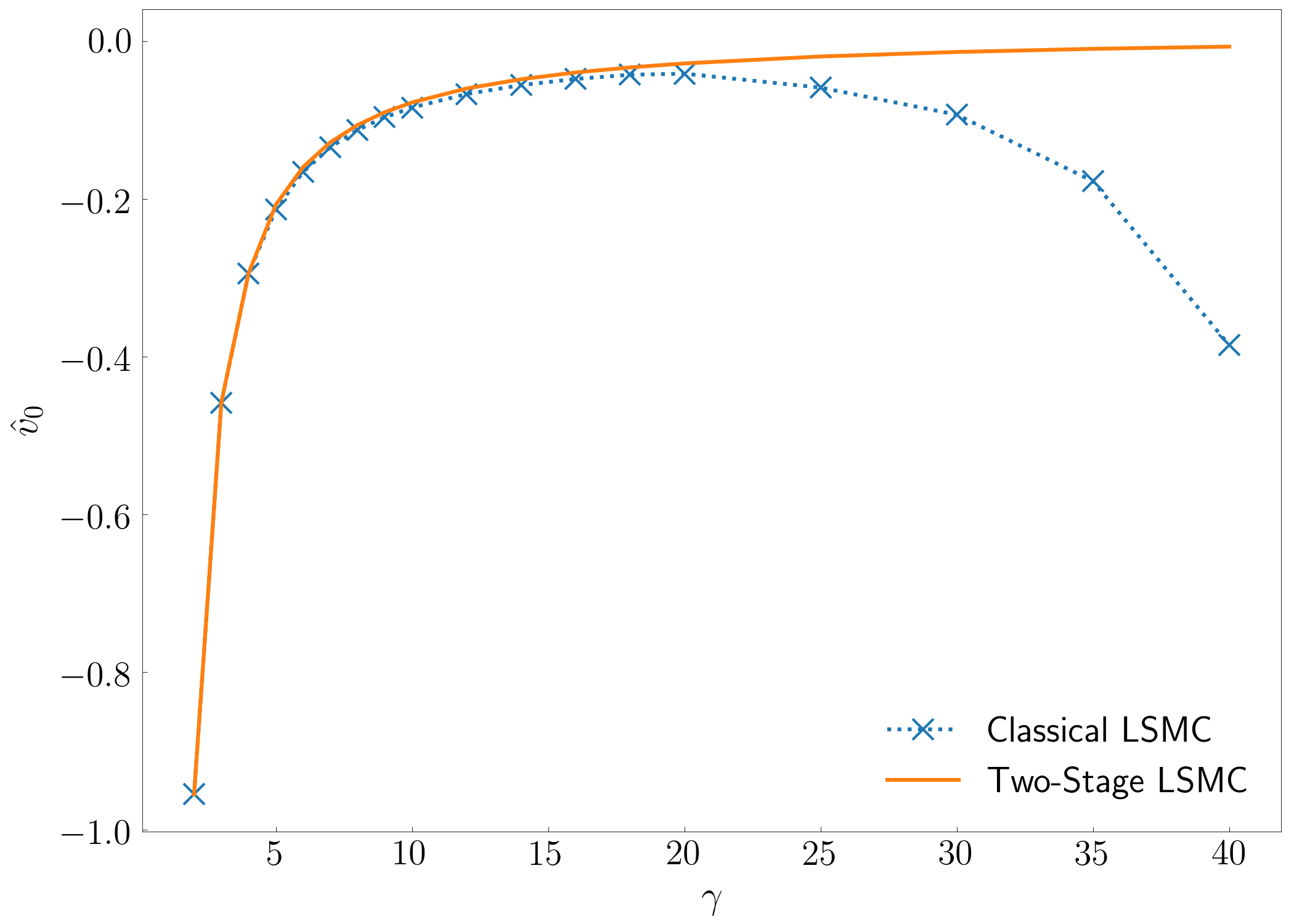}
\par\end{center}%
\end{minipage}
\end{figure}

The following experiment aims at validating the two-stage LSMC method
via a comparison to the classical LSMC method. We first study a CRRA
utility optimization example. It has been noted that a simulation-and-regression
approach can generate large numerical errors when the utility function
is highly nonlinear (high risk aversion), see for example \citet{vanBinsbergen2007},
\citet{Garlappi2009} and \citet{Denault2017}. We apply the two-stage
LSMC method and the classical LSMC method to CRRA utility optimization,
and then compare the resulting initial value function estimates $\hat{v}_{0}=\frac{1}{M}\sum_{m=1}^{M}(\hat{W}_{t_{N}})^{1-\gamma}/(1-\gamma)$
for a one-year time horizon with monthly rebalancing. Following \citet{Zhang2018},
we choose $M=10,000$ sample paths to ensure numerical stability of
the solution. For the classical LSMC method, we include the utility
function itself as part of the regression basis, so that the regression
basis can be adjusted to some extent to the risk-aversion parameter.
Figure \ref{fig:LSMC-2LSMC-CRRA} shows that the classical LSMC method
becomes unstable when the value of $\gamma$ is high, while the two-stage
LSMC method converges quite well. In our experiment, the two-stage
LSMC method can approximate the CRRA utility optimization approach
well up to $\gamma=100$.

We then compare our two-stage LSMC to the classical LSMC for solving
the STRS. To check the possibility of heteroskedastic residuals, we
calibrate a state-dependent standard deviation $\sigma\left(z,w\right)$
as described in Subsection \ref{subsec:State-dependent-SD} and compare
it with the original two-stage LSMC method in which the standard deviation
only depends on the portfolio decision. In particular, we use a simple
linear basis to approximate the logarithmic standard deviation. Figure
\ref{tab:LSMC-2LSMC-STRS} shows that the two-stage LSMC method substantially
improves the estimates $\hat{v}_{0}$ and the return distributions,
compared to the classical LSMC approach, while using a state-dependent
standard deviation does not significantly improve the results, suggesting
that the assumption of homoskedastic residuals is reasonable.

\begin{table}[H]
\caption{Two-stage LSMC v.s. classical LSMC for STRS\label{tab:LSMC-2LSMC-STRS}}

\setlength\tabcolsep{2.5pt}

\begin{minipage}[t]{1\columnwidth}%
\begin{center}
{\footnotesize{}}%
\begin{tabular}{ccccccccccccccccc}
 &  &  & \multicolumn{4}{l}{{\footnotesize{}Classical LSMC}} &  & \multicolumn{4}{l}{{\footnotesize{}Two-Stage LSMC}} &  & \multicolumn{4}{l}{{\footnotesize{}Two-Stage LSMC + $\sigma\!\left(z,\!w\right)$}}\tabularnewline
\cline{4-17} \cline{5-17} \cline{6-17} \cline{7-17} \cline{8-17} \cline{9-17} \cline{10-17} \cline{11-17} \cline{12-17} \cline{13-17} \cline{14-17} \cline{15-17} \cline{16-17} \cline{17-17} 
{\footnotesize{}$L_{{\scriptscriptstyle \!W}}$} & {\footnotesize{}$U_{\!{\scriptscriptstyle W}}$} &  & {\footnotesize{}$\hat{v}_{0}$} & {\footnotesize{}$\mathbb{E}\!\left[W_{T}\right]$ } & {\footnotesize{}$\text{SD}\!\left[W_{T}\right]$} & {\footnotesize{}$\mathbb{P}\!\left[W_{T}\!\!<\!\!1\right]$} &  & {\footnotesize{}$\hat{v}_{0}$} & {\footnotesize{}$\mathbb{E}\!\left[W_{T}\right]$ } & {\footnotesize{}$\text{SD}\!\left[W_{T}\right]$} & {\footnotesize{}$\mathbb{P}\!\left[W_{T}\!\!<\!\!1\right]$} &  & {\footnotesize{}$\hat{v}_{0}$} & {\footnotesize{}$\mathbb{E}\!\left[W_{T}\right]$ } & {\footnotesize{}$\text{SD}\!\left[W_{T}\right]$} & {\footnotesize{}$\mathbb{P}\!\left[W_{T}\!\!<\!\!1\right]$}\tabularnewline
\hline 
{\footnotesize{}1} & {\footnotesize{}1.1} &  & {\footnotesize{}0.0058} & {\footnotesize{}1.1571} & {\footnotesize{}0.1847} & {\footnotesize{}0.1244} &  & {\footnotesize{}0.0574} & {\footnotesize{}1.0596} & {\footnotesize{}0.0272} & {\footnotesize{}0.0028} &  & {\footnotesize{}0.0475} & {\footnotesize{}1.0499} & {\footnotesize{}0.0318} & {\footnotesize{}0.0095}\tabularnewline
{\footnotesize{}1} & {\footnotesize{}1.2} &  & {\footnotesize{}0.0292} & {\footnotesize{}1.1609} & {\footnotesize{}0.1709} & {\footnotesize{}0.1077} &  & {\footnotesize{}0.0922} & {\footnotesize{}1.0883} & {\footnotesize{}0.0405} & {\footnotesize{}0.0128} &  & {\footnotesize{}0.0904} & {\footnotesize{}1.0867} & {\footnotesize{}0.0405} & {\footnotesize{}0.0122}\tabularnewline
{\footnotesize{}1} & {\footnotesize{}1.3} &  & {\footnotesize{}0.0608} & {\footnotesize{}1.1631} & {\footnotesize{}0.1542} & {\footnotesize{}0.0832} &  & {\footnotesize{}0.1190} & {\footnotesize{}1.1126} & {\footnotesize{}0.0588} & {\footnotesize{}0.0178} &  & {\footnotesize{}0.1239} & {\footnotesize{}1.1164} & {\footnotesize{}0.0609} & {\footnotesize{}0.0192}\tabularnewline
{\footnotesize{}1} & {\footnotesize{}1.4} &  & {\footnotesize{}0.0918} & {\footnotesize{}1.1663} & {\footnotesize{}0.1597} & {\footnotesize{}0.0656} &  & {\footnotesize{}0.1393} & {\footnotesize{}1.1296} & {\footnotesize{}0.0832} & {\footnotesize{}0.0244} &  & {\footnotesize{}0.1446} & {\footnotesize{}1.1351} & {\footnotesize{}0.0893} & {\footnotesize{}0.0286}\tabularnewline
{\footnotesize{}1} & {\footnotesize{}1.5} &  & {\footnotesize{}0.1199} & {\footnotesize{}1.1692} & {\footnotesize{}0.1625} & {\footnotesize{}0.0503} &  & {\footnotesize{}0.1578} & {\footnotesize{}1.1449} & {\footnotesize{}0.1078} & {\footnotesize{}0.0299} &  & {\footnotesize{}0.1596} & {\footnotesize{}1.1491} & {\footnotesize{}0.1165} & {\footnotesize{}0.0321}\tabularnewline
{\footnotesize{}1} & {\footnotesize{}1.6} &  & {\footnotesize{}0.1455} & {\footnotesize{}1.1721} & {\footnotesize{}0.1641} & {\footnotesize{}0.0454} &  & {\footnotesize{}0.1718} & {\footnotesize{}1.1563} & {\footnotesize{}0.1264} & {\footnotesize{}0.0352} &  & {\footnotesize{}0.1728} & {\footnotesize{}1.1596} & {\footnotesize{}0.1359} & {\footnotesize{}0.0413}\tabularnewline
{\footnotesize{}1} & {\footnotesize{}$\infty$} &  & {\footnotesize{}0.1903} & {\footnotesize{}1.1743} & {\footnotesize{}0.1652} & {\footnotesize{}0.0483} &  & {\footnotesize{}0.1934} & {\footnotesize{}1.1684} & {\footnotesize{}0.1635} & {\footnotesize{}0.0423} &  & {\footnotesize{}0.1938} & {\footnotesize{}1.1688} & {\footnotesize{}0.1625} & {\footnotesize{}0.0446}\tabularnewline
\hline 
\end{tabular}
\par\end{center}%
\end{minipage}
\end{table}

\subsection{STRS and CRRA}

\begin{figure}[H]
\begin{centering}
\caption{Terminal wealth distribution: comparison between STRS and CRRA\label{fig:benchmark}}
\par\end{centering}
\smallskip
\centering{}%
\begin{minipage}[t]{0.4\columnwidth}%
\includegraphics[scale=0.55]{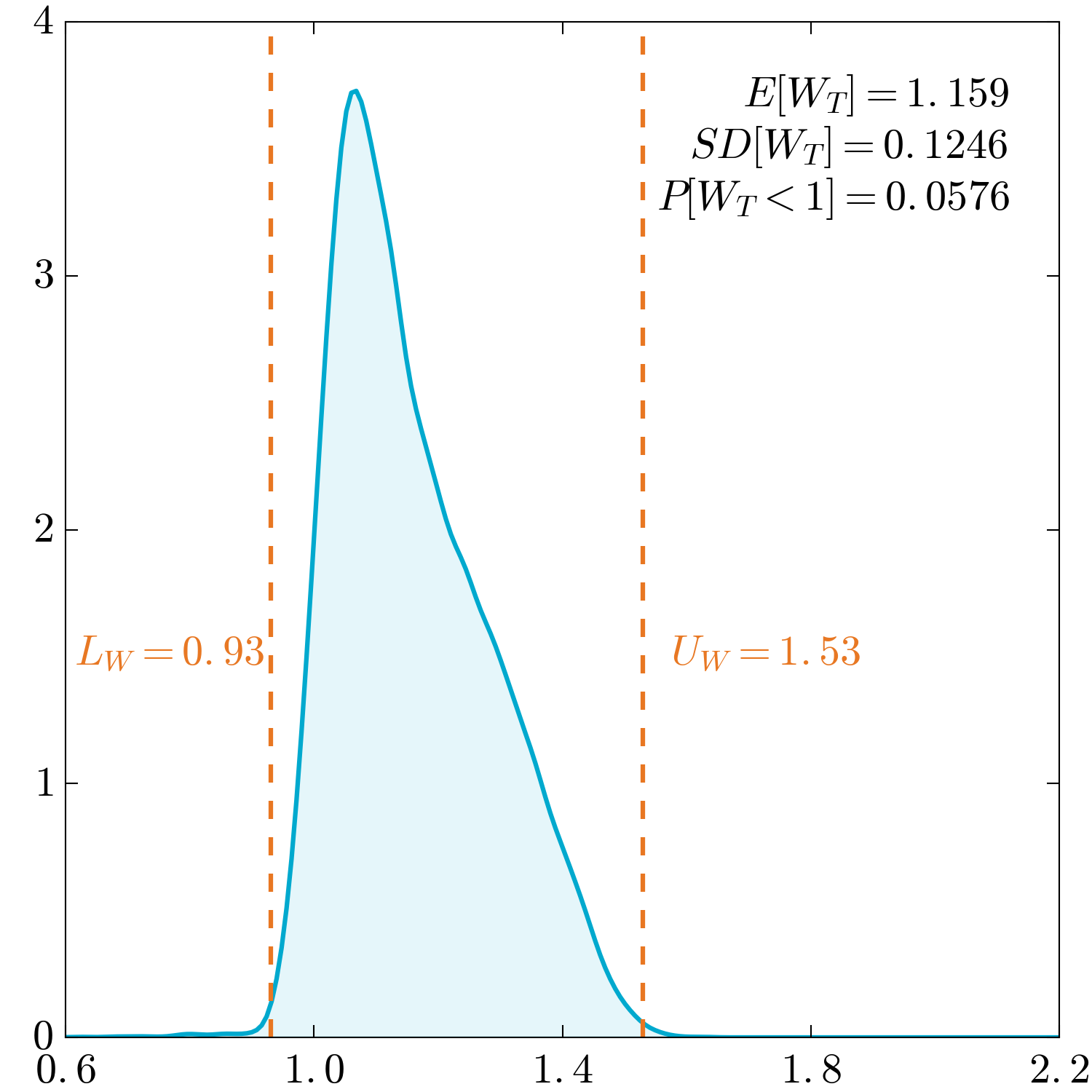}%
\end{minipage}\qquad{}%
\begin{minipage}[t]{0.4\columnwidth}%
\includegraphics[scale=0.55]{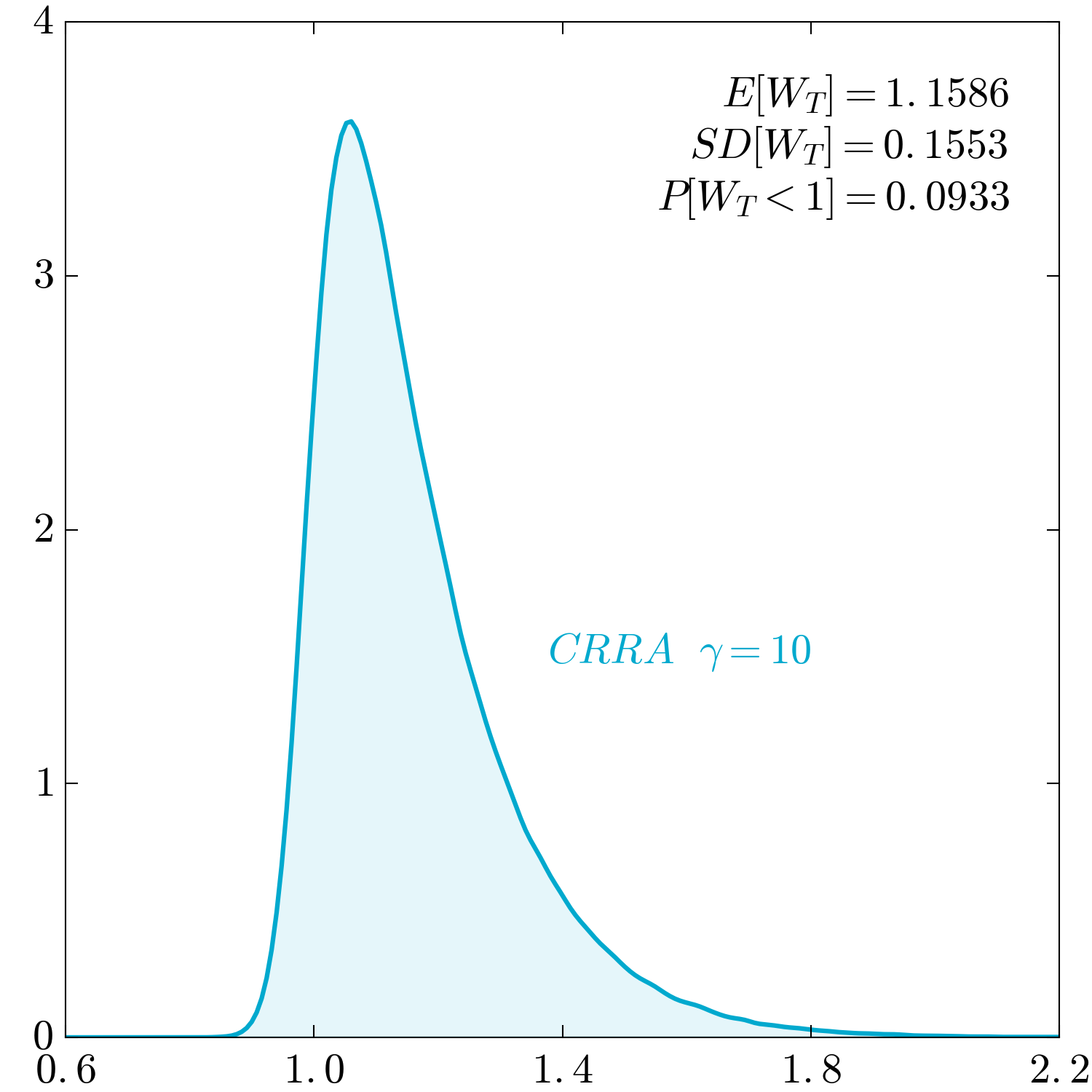}%
\end{minipage}
\end{figure}

We now compare the STRS to the CRRA utility optimization approach.
Our main finding regarding this comparison is that for each risk aversion
level $\gamma$ of the CRRA utility approach, one can find a target
range $[L_{{\scriptscriptstyle \!W}},U_{{\scriptscriptstyle \!W}}]$
such that the STRS delivers a similar expectation, but with a lower
standard deviation and a lower downside risk. As an illustration,
Figure \ref{fig:benchmark} shows how the STRS with $[L_{{\scriptscriptstyle \!W}},U_{{\scriptscriptstyle \!W}}]=[0.93,1.53]$
outperforms the CRRA utility approach with $\gamma=10$. Despite the
better statistical moments of the STRS, the shorter right tail of
the STRS compared to the CRRA utility approach can be deemed a shortcoming
of our approach, though giving up some upside potential is the reason
for the improved downside risk protection compared to the CRRA utility
approach.

\begin{figure}[H]
\begin{centering}
\caption{Comparison with CRRA: risk-return trade-off\label{fig:frontier}}
\par\end{centering}
\smallskip
\centering{}%
\begin{minipage}[t]{0.4\columnwidth}%
\includegraphics[scale=0.13]{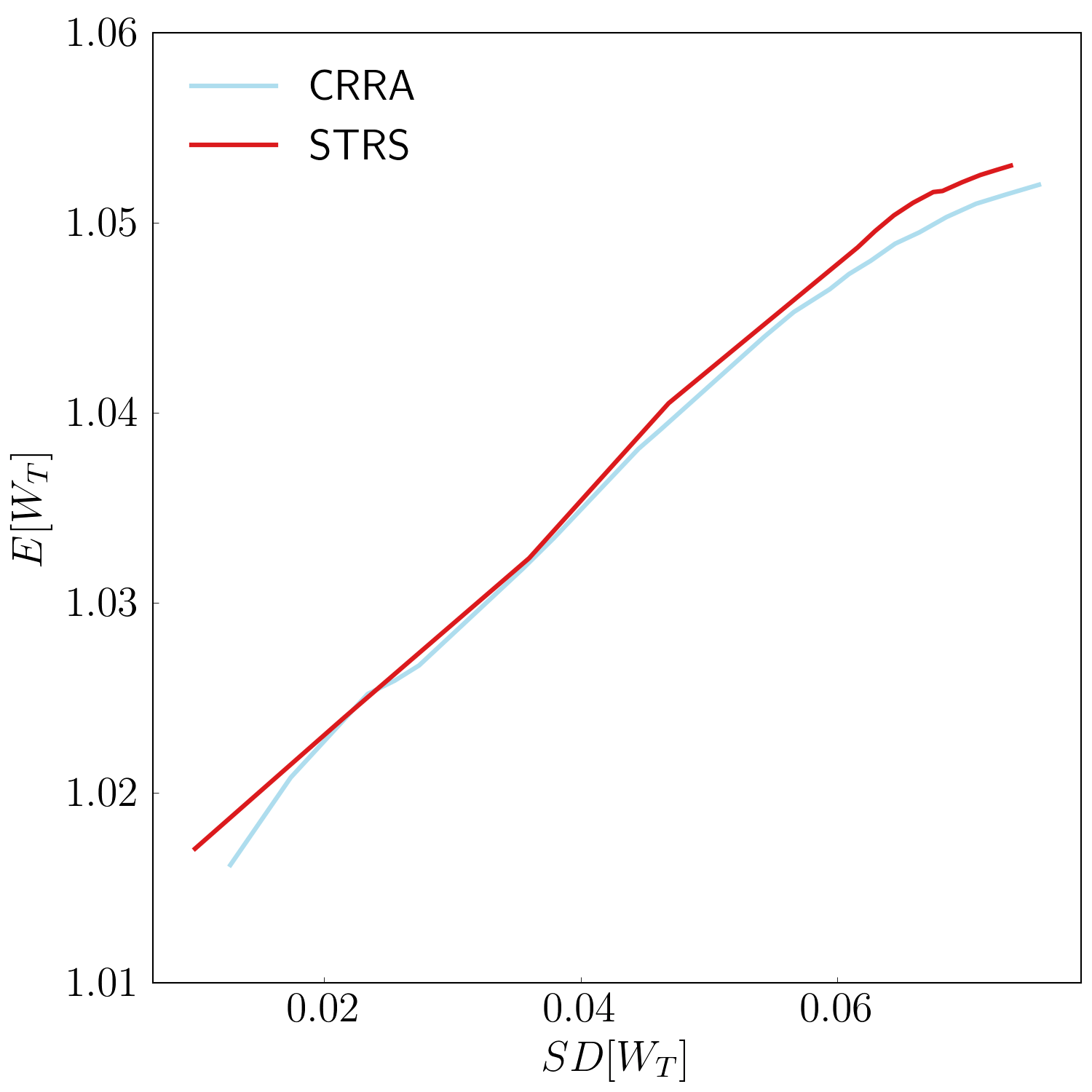}%
\end{minipage}\qquad{}%
\begin{minipage}[t]{0.4\columnwidth}%
\includegraphics[scale=0.13]{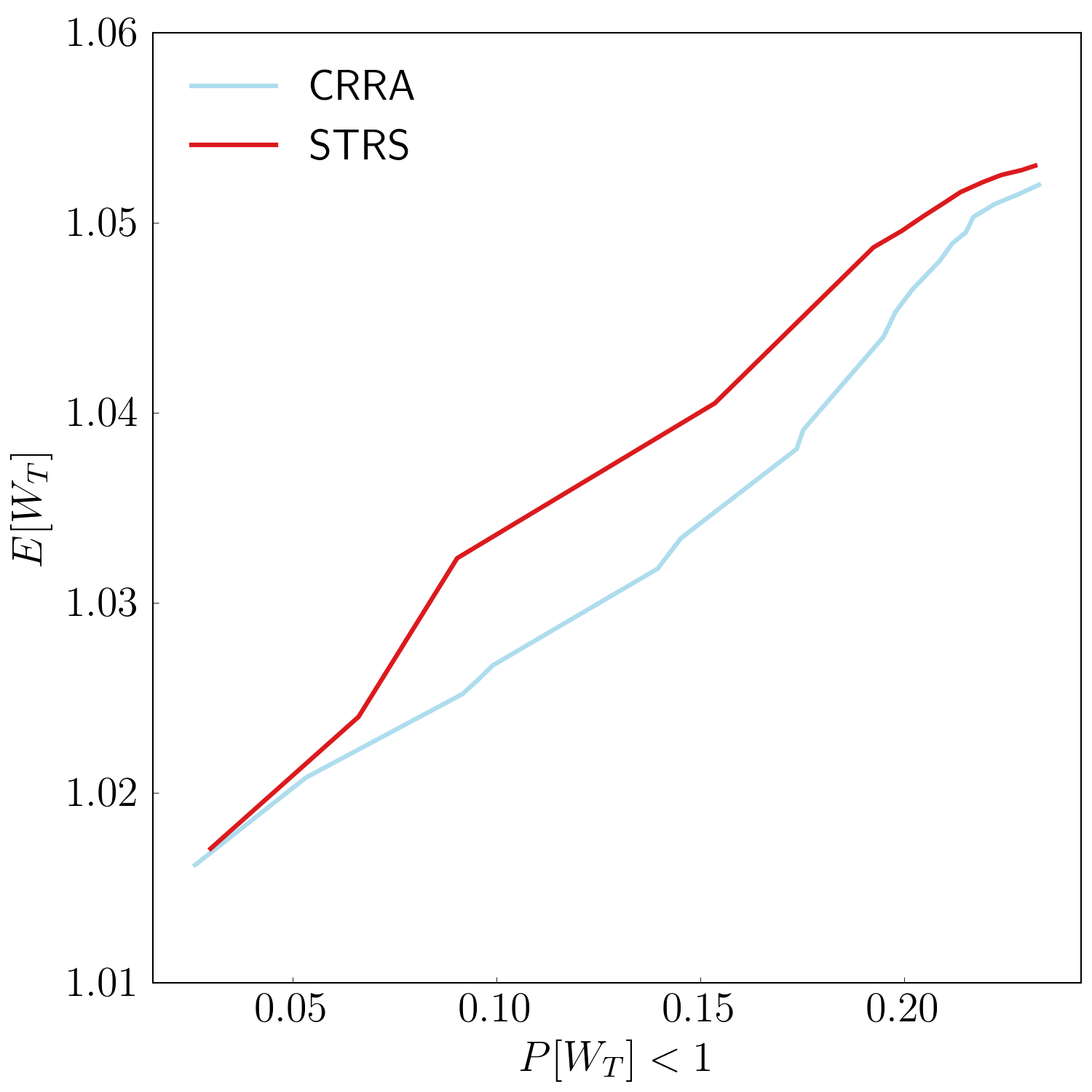}%
\end{minipage}
\end{figure}

To provide a more comprehensive comparison, we now report two risk-return
trade-offs: the mean-variance efficient frontier and the trade-off
between return and downside risk. Figure \ref{fig:frontier} displays
the efficient frontiers of the STRS (for different combinations of
$L_{{\scriptscriptstyle \!W}}$ and $U_{{\scriptscriptstyle \!W}}$)
and the CRRA utility approach (for different $\gamma$ levels) for
a three-month investment horizon. The results show that the STRS and
the CRRA utility approach trace out a similar mean-variance efficient
frontier, while the STRS delivers a better downside risk-return trade-off.
Remark that the STRS and the CRRA utility approach produce similar
results when the risk-aversion parameter is either very small (risk-neutral)
or very high, while the STRS is preferable for intermediate risk-aversion
levels.

A theoretical proof of the higher efficiency of the STRS over the
classical utility strategies would be desirable to corroborate our
numerical findings. However, given for example the difficulty in deriving
an explicit optimal allocation for a single trading period with a
simpler downside risk minimization objective (\citealt{Klebaner2017}),
a theoretical proof of the higher efficiency of the STRS over the
classical utility strategies might be out of reach. We thus leave
this question for further research.

\subsection{Extensions}

\begin{figure}[H]
\caption{Terminal wealth distributions using FTRS\label{fig:prob}}

\smallskip
\begin{centering}
\begin{minipage}[t]{0.4\columnwidth}%
\includegraphics[scale=0.55]{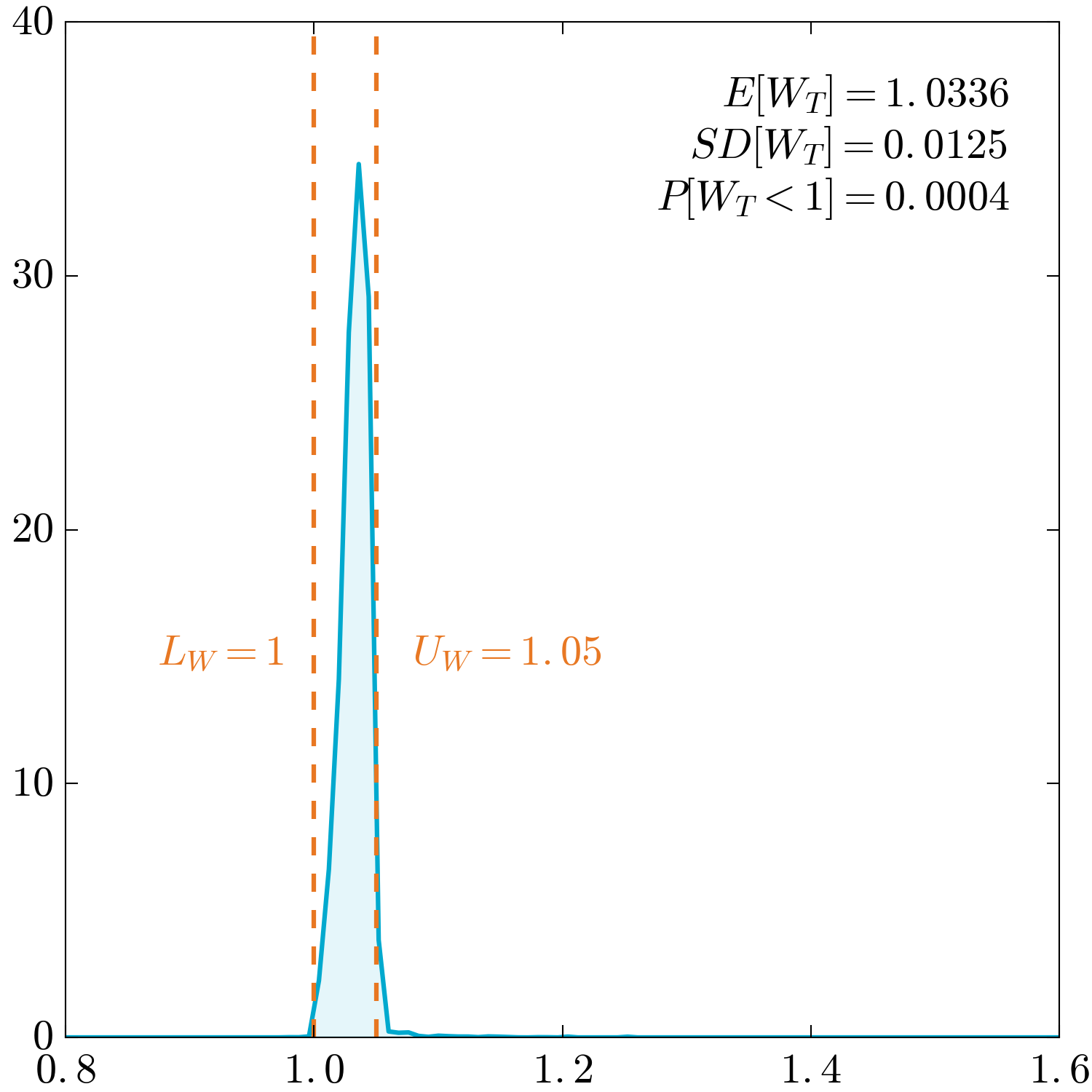}%
\end{minipage}\qquad{}%
\begin{minipage}[t]{0.4\columnwidth}%
\includegraphics[scale=0.55]{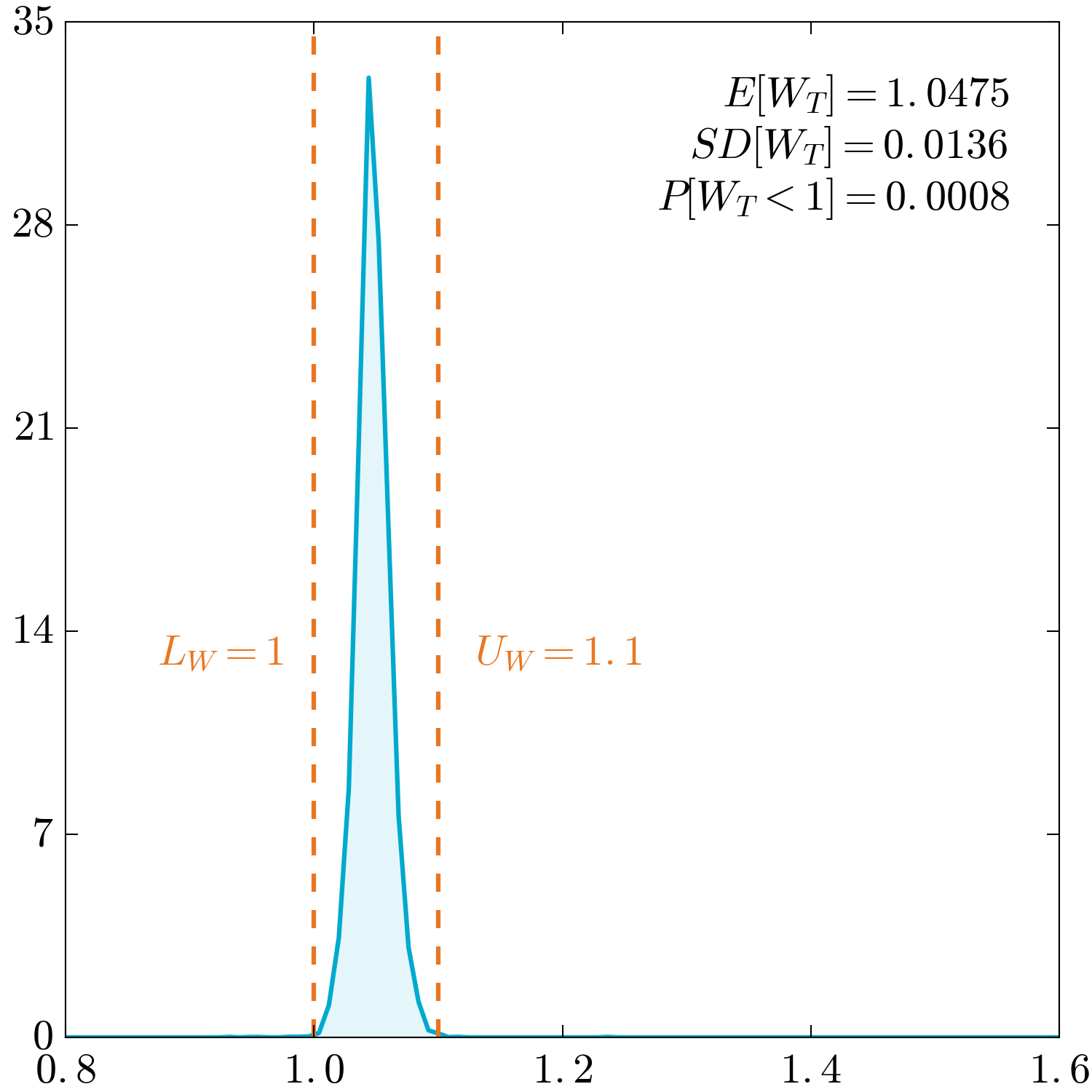}%
\end{minipage}
\par\end{centering}
\centering{}%
\begin{minipage}[t]{0.4\columnwidth}%
\includegraphics[scale=0.55]{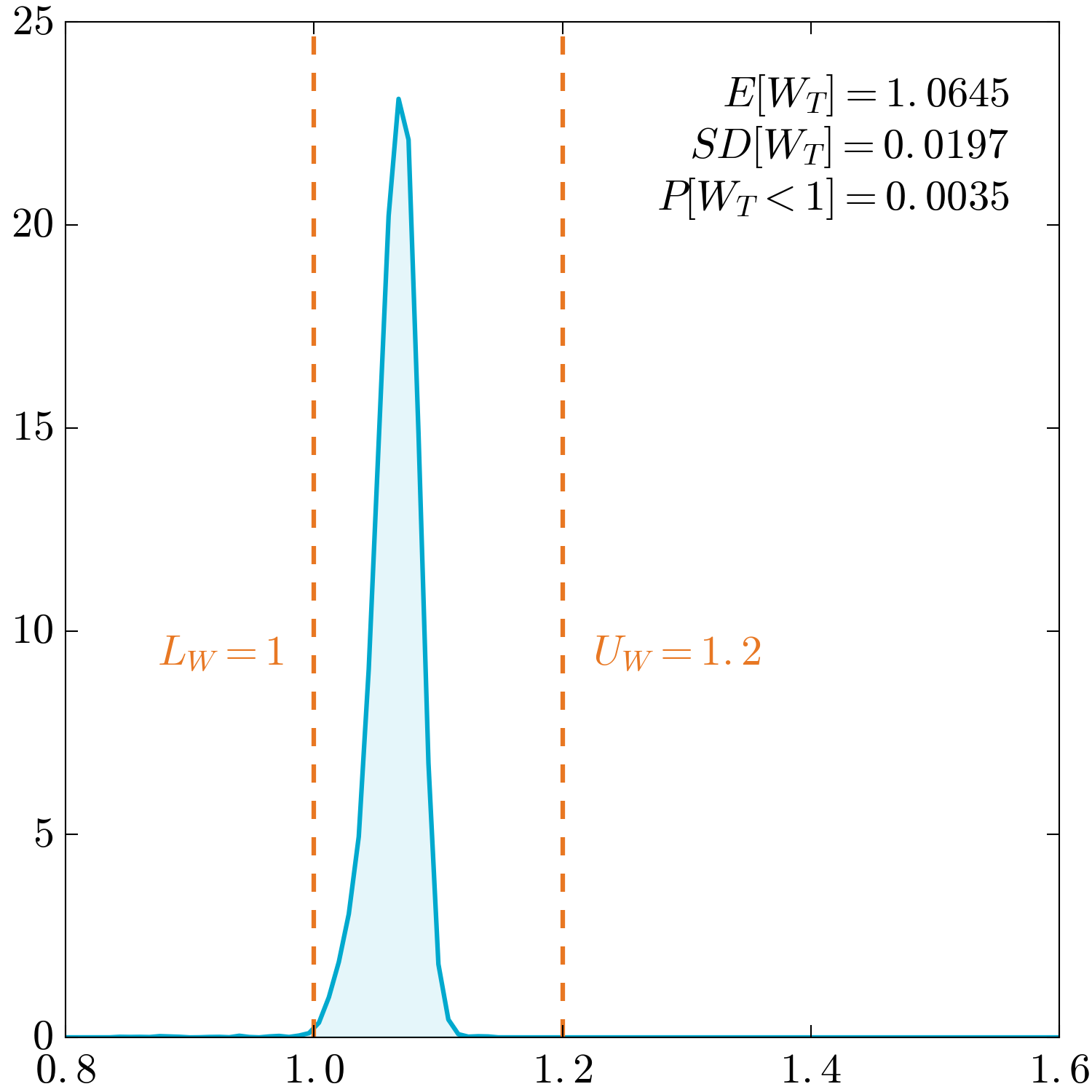}%
\end{minipage}\qquad{}%
\begin{minipage}[t]{0.4\columnwidth}%
\includegraphics[scale=0.55]{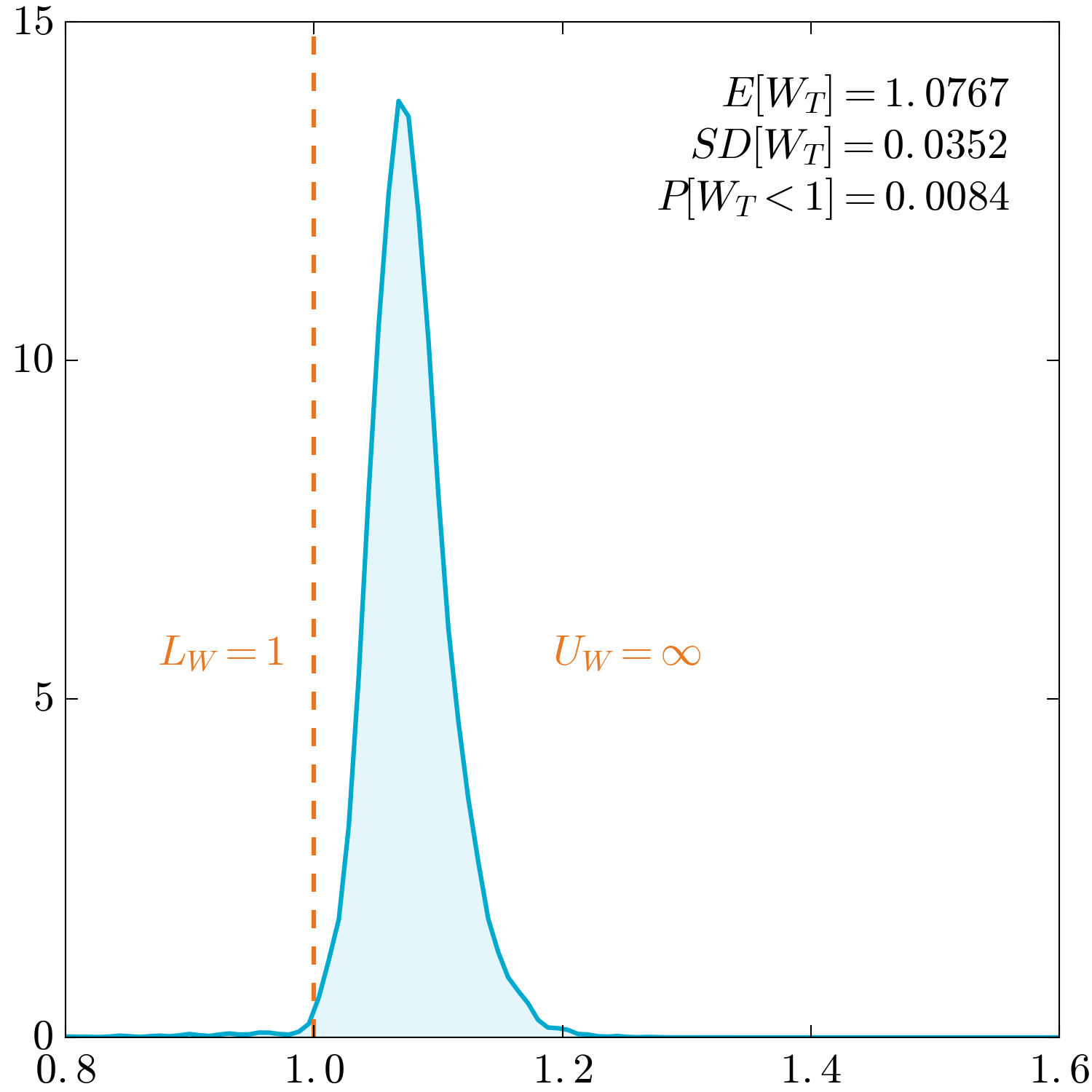}%
\end{minipage}
\end{figure}

This subsection discusses the wealth distributions produced by the
modified target range strategies described in Section \ref{sec:Extensions}.
Figure \ref{fig:prob} provides examples for the flat target range
strategy (FTRS) with $L_{\!{\scriptscriptstyle W}}=1.0$ and $U_{\!{\scriptscriptstyle W}}=1.05$,
$1.10$, $1.20$ and $+\infty$. The main observation is that, as
expected, the probability of the terminal wealth lying outside the
predefined range $[L_{{\scriptscriptstyle \!W}},U_{{\scriptscriptstyle \!W}}]$
is smaller than for the STRS (refer to Figure \ref{fig:target} for
comparison). This is the main strength of the FTRS: downside risk
is kept to a minimum, while the price to pay for this safety is the
inability to generate high returns. Finally, the wealth distribution
is less sensitive to the choice of $U_{\!{\scriptscriptstyle W}}$:
the distribution is tight even when $U_{\!{\scriptscriptstyle W}}=\infty$,
given the absence of incentive to chase high returns. 

In theory, if one wants to maximize the probability that the terminal
wealth lies within the targeted range with the lower bound $L_{\!{\scriptscriptstyle W}}=1.0$
and a large enough upper bound $U_{\!{\scriptscriptstyle W}}$, the
optimal decision should be to allocate all the capital to the risk-free
asset. Numerically though, it is difficult to guarantee a full allocation
in the risk-free asset at all times and for all paths. Intuitively,
the reason for this is the following: for the portfolios allocated
mostly to the risk-free asset, most, if not all, of the terminal wealth
realizations will lie within the targeted range, which makes the value
function flat and almost invariant among these convervative portfolio
allocations.

\begin{figure}[H]
\begin{centering}
\caption{Excess terminal wealth distributions with relative target range strategies\label{fig:ex_return}}
\par\end{centering}
\smallskip
\begin{centering}
\begin{minipage}[t]{0.4\columnwidth}%
\includegraphics[scale=0.55]{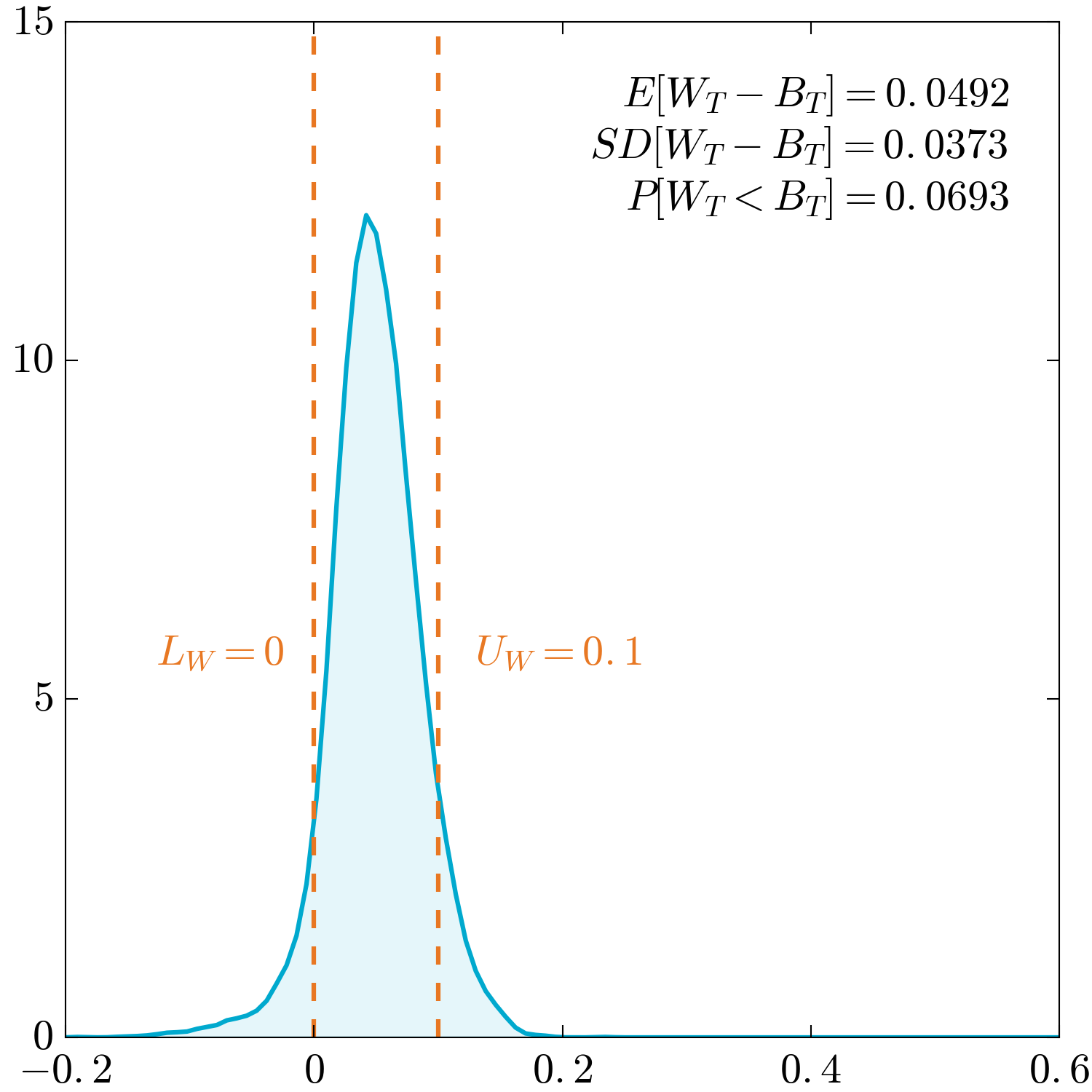}%
\end{minipage}\qquad{}%
\begin{minipage}[t]{0.4\columnwidth}%
\includegraphics[scale=0.55]{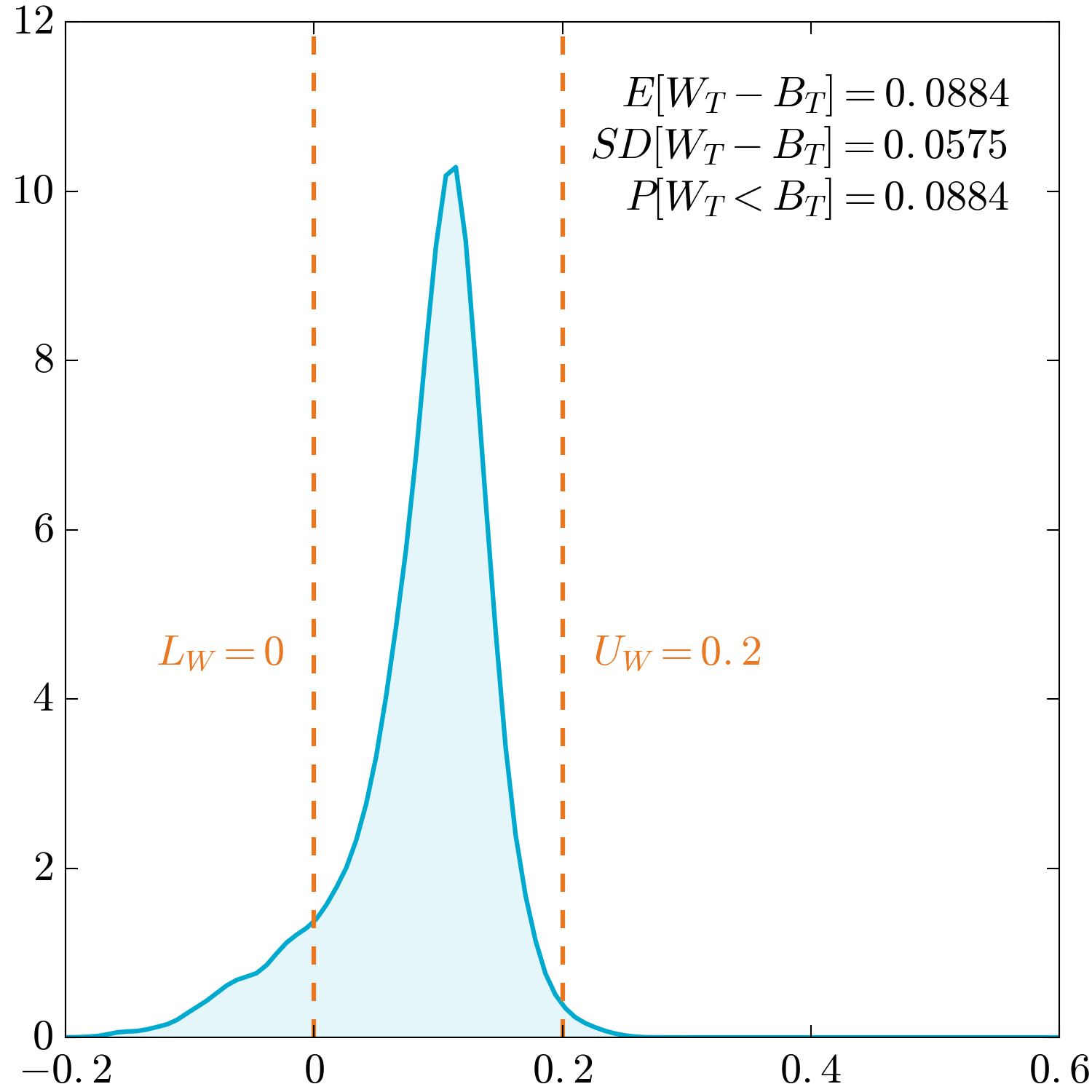}%
\end{minipage}
\par\end{centering}
\begin{centering}
\begin{minipage}[t]{0.4\columnwidth}%
\includegraphics[scale=0.55]{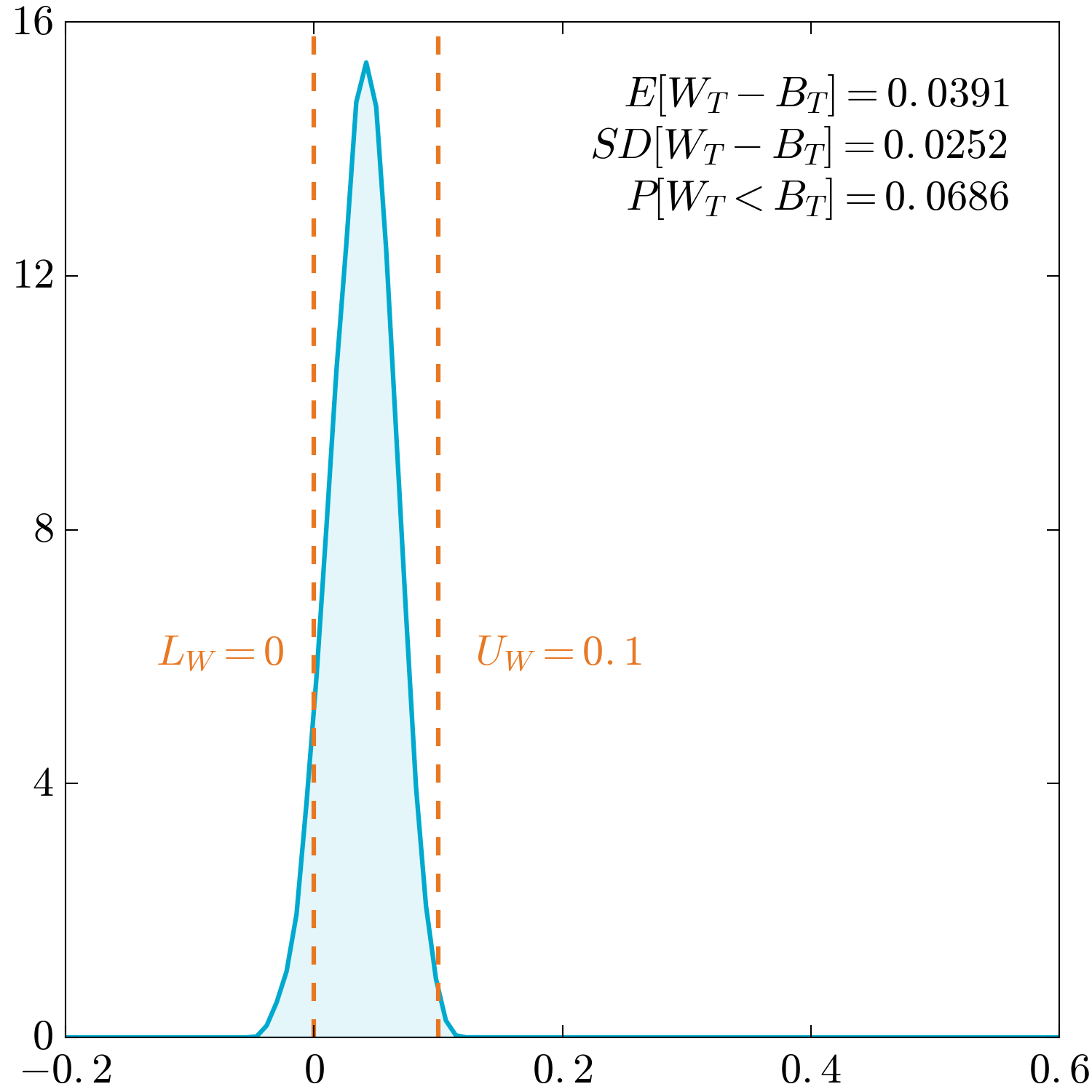}%
\end{minipage}\qquad{}%
\begin{minipage}[t]{0.4\columnwidth}%
\includegraphics[scale=0.55]{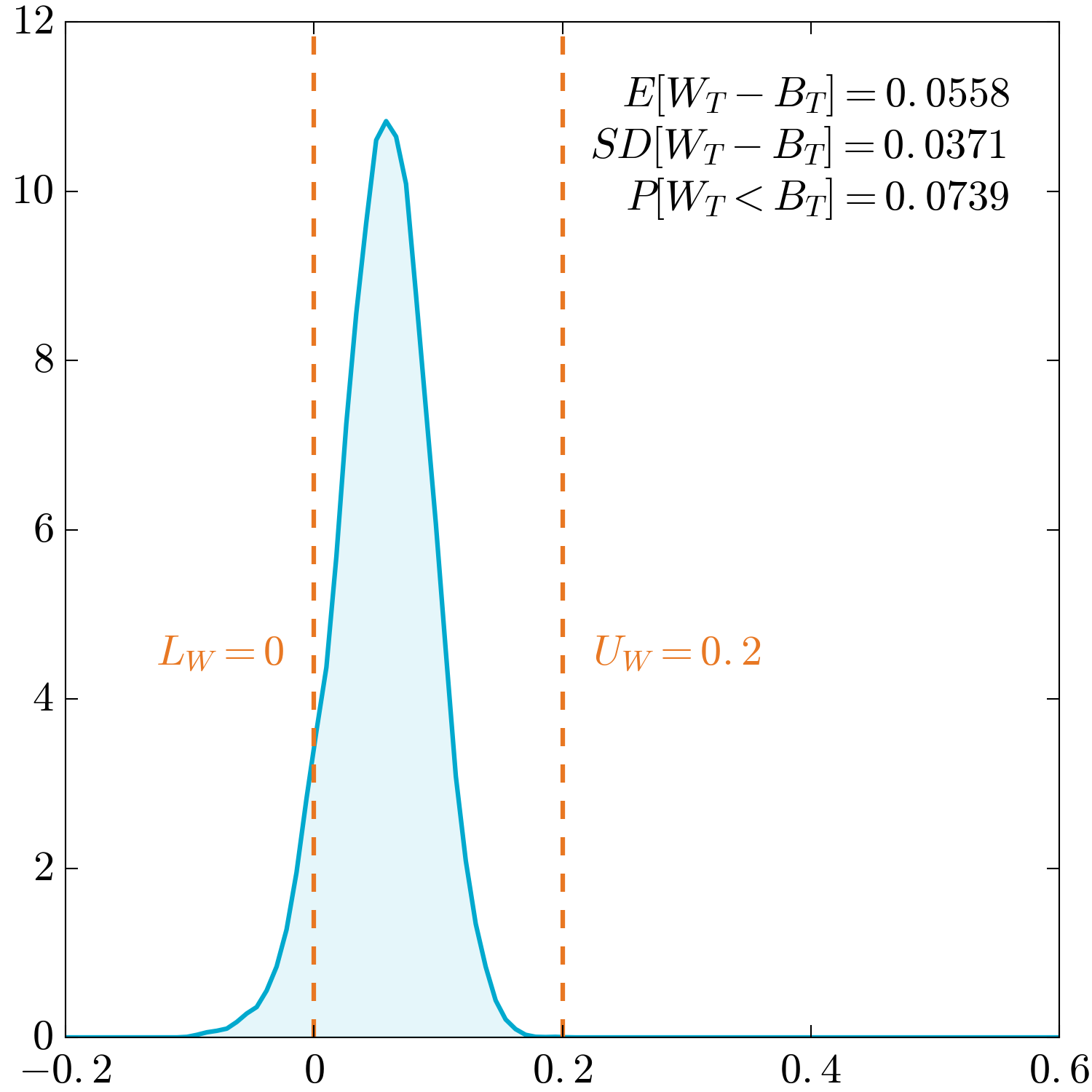}%
\end{minipage}
\par\end{centering}
\centering{}{\footnotesize{}Excess wealth distributions of STRS (top
row) and FTRS (bottom row)}{\footnotesize\par}
\end{figure}

Figure \ref{fig:ex_return} provides some examples for the relative
target range strategy (RTRS) with a passive equal-weight portfolio
as benchmark. The probability that the portfolio value underperforms
the benchmark portfolio remains small (around $6\%-8\%$ for the excess
return distributions), though higher than those provided by absolute
targets. The reason for this is that the passive equal-weight benchmark
already delivers a high expected return, therefore outperforming it
requires taking more risk than what was necessary in the previous
absolute return target examples.

\section{Conclusions\label{sec:Conclusion}}

This paper introduces the skewed target range strategy (STRS) for
portfolio optimization problems. The STRS maximizes the expected portfolio
value while simultaneously restraining the bulk of the return distribution
within a predefined range. This joint goal is achieved with an unconstrained
optimization formulation, which achieves, in a simpler manner, similar
results to those that can be expected from more complex constrained
optimization methods. To illustrate the effectiveness of the STRS,
we study a multiperiod portfolio optimization problem and propose
a two-stage least squares Monte Carlo (LSMC) method to handle the
new objective function. The two-stage regression method can also be
adopted for general investment objectives such as the smooth constant
relative risk aversion (CRRA) utility. We show that this regression
method substantially improves the numerical stability of the LSMC
algorithm compared to direct regression. We show that the STRS achieves
a similar mean-variance efficient frontier while delivering a better
downside risk-return trade-off, compared to the CRRA utility approach.
We find that the recommended level for the lower bound of the target
range is the initial portfolio value, at which the standard deviation
and the downside risk of the terminal portfolio value are marginally
minimized. From there, the upper bound of the target range can be
set based on risk preferences.

Going further, the unconstrained optimization formulation used by
the STRS, built upon an indicator function, has the potential to incorporate
additional range constraints on other dynamic risk measures such as
realized volatility or maximum drawdown. This is an area we wish to
investigate in future research.

\paragraph*{Acknowledgments }

The authors are grateful to Dr. Wen Chen and the two anonymous referees
for their valuable comments and remarks.\bibliographystyle{chicago}
\bibliography{bib_objective,bib_portfolio}

\begin{thebibliography}{}

\bibitem[\protect\citeauthoryear{Agarwal and Naik}{Agarwal and
  Naik}{2004}]{Agarwal2004}
Agarwal, V. and N.~Y. Naik (2004).
\newblock Risks and portfolio decisions involving hedge funds.
\newblock {\em Review of Financial Studies\/}~{\em 17\/}(1), 63--98.

\bibitem[\protect\citeauthoryear{Alexander and Baptista}{Alexander and
  Baptista}{2002}]{Alexander2002}
Alexander, G.~J. and A.~M. Baptista (2002).
\newblock Economic implications of using a mean-{V}a{R} model for portfolio
  selection: A comparison with mean-variance analysis.
\newblock {\em Journal of Economic Dynamics and Control\/}~{\em 26\/}(7-8),
  1159--1193.

\bibitem[\protect\citeauthoryear{Andreasson and Shevchenko}{Andreasson and
  Shevchenko}{2018}]{Andreasson2018}
Andreasson, J. and P.~Shevchenko (2018).
\newblock Bias-corrected least-squares {M}onte {C}arlo for utility based
  optimal stochastic control problems.
\newblock SSRN:2985828.

\bibitem[\protect\citeauthoryear{Balata and Palczewski}{Balata and
  Palczewski}{2018}]{Balata18}
Balata, A. and J.~Palczewski (2018).
\newblock Regress-{L}ater {M}onte {C}arlo for optimal control of {M}arkov
  processes.

\bibitem[\protect\citeauthoryear{Barberis}{Barberis}{2012}]{Barberis2012}
Barberis, N. (2012).
\newblock A model of casino gambling.
\newblock {\em Management Science\/}~{\em 58\/}(1), 35--51.

\bibitem[\protect\citeauthoryear{Brandt, Goyal, Santa-Clara, and Stroud}{Brandt
  et~al.}{2005}]{Brandt2005}
Brandt, M., A.~Goyal, P.~Santa-Clara, and J.~Stroud (2005).
\newblock A simulation approach to dynamic portfolio choice with an application
  to learning about return predictability.
\newblock {\em Review of Financial Studies\/}~{\em 18}, 831--873.

\bibitem[\protect\citeauthoryear{Brogan and Stidham~Jr.}{Brogan and
  Stidham~Jr.}{2005}]{Brogan2005}
Brogan, A.~J. and S.~Stidham~Jr. (2005).
\newblock A note on separation in mean-lower-partial-moment portfolio
  optimization with fixed and moving targets.
\newblock {\em IIE Transactions\/}~{\em 37\/}(10), 901--906.

\bibitem[\protect\citeauthoryear{Browne}{Browne}{1999a}]{Browne1999a}
Browne, S. (1999a).
\newblock Beating a moving target: Optimal portfolio strategies for
  outperforming a stochastic benchmark.
\newblock {\em Finance and Stochastics\/}~{\em 3\/}(3), 275--294.

\bibitem[\protect\citeauthoryear{Browne}{Browne}{1999b}]{Browne1999b}
Browne, S. (1999b).
\newblock The risk and rewards of minimizing shortfall probability.
\newblock {\em Journal of Portfolio Management\/}~{\em 25\/}(4), 76--85.

\bibitem[\protect\citeauthoryear{Carriere}{Carriere}{1996}]{Carriere1996}
Carriere, J. (1996).
\newblock Valuation of the early-exercise price for options using simulations
  and nonparametric regression.
\newblock {\em Insurance: Mathematics and Economics\/}~{\em 19\/}(1), 19--30.

\bibitem[\protect\citeauthoryear{Cong and Oosterlee}{Cong and
  Oosterlee}{2016a}]{Cong2016}
Cong, F. and C.~W. Oosterlee (2016a).
\newblock Multi-period mean-variance portfolio optimization based on {M}onte
  {Carlo} simulation.
\newblock {\em Journal of Economic Dynamics and Control\/}~{\em 64}, 23--38.

\bibitem[\protect\citeauthoryear{Cong and Oosterlee}{Cong and
  Oosterlee}{2016b}]{Cong2016b}
Cong, F. and C.~W. Oosterlee (2016b).
\newblock On pre-commitment aspects of a time-consistent strategy for a
  mean-variance investor.
\newblock {\em Journal of Economic Dynamics and Control\/}~{\em 70\/}(1),
  178--193.

\bibitem[\protect\citeauthoryear{Cong and Oosterlee}{Cong and
  Oosterlee}{2017}]{Cong2017}
Cong, F. and C.~W. Oosterlee (2017).
\newblock Accurate and robust numerical methods for the dynamic portfolio
  management problem.
\newblock {\em Computational Economics\/}~{\em 49\/}(3), 433--458.

\bibitem[\protect\citeauthoryear{Dang, Forsyth, and Vetzal}{Dang
  et~al.}{2017}]{Dang2017}
Dang, D.-M., P.~Forsyth, and K.~Vetzal (2017).
\newblock The 4\% strategy revisited: a pre-commitment mean-variance optimal
  approach to wealth management.
\newblock {\em Quantitative Finance\/}~{\em 17\/}(3), 335--351.

\bibitem[\protect\citeauthoryear{Davis and Norman}{Davis and
  Norman}{1990}]{Davis1990}
Davis, M. and A.~Norman (1990).
\newblock Portfolio selection with transaction costs.
\newblock {\em Mathematics of Operations Research\/}~{\em 15\/}(4), 676--713.

\bibitem[\protect\citeauthoryear{Denault and Simonato}{Denault and
  Simonato}{2017}]{Denault2017}
Denault, M. and J.-G. Simonato (2017).
\newblock Dynamic portfolio choices by simulation-and-regression: Revisiting
  the issue of value function vs portfolio weight recursions.
\newblock {\em Computers and Operations Reseach\/}~{\em 79}, 174--189.

\bibitem[\protect\citeauthoryear{Franks}{Franks}{1992}]{Franks1992}
Franks, E.~C. (1992).
\newblock Targeting excess-of-benchmark returns.
\newblock {\em Journal of Portfolio Management\/}~{\em 18\/}(4), 6--12.

\bibitem[\protect\citeauthoryear{Gaivoronski, Krylov, and van~der
  Wijst}{Gaivoronski et~al.}{2005}]{Gaivoronski2005}
Gaivoronski, A.~A., S.~Krylov, and N.~van~der Wijst (2005).
\newblock Optimal portfolio selection and dynamic benchmark tracking.
\newblock {\em European Journal of Operational Research\/}~{\em 163\/}(1),
  115--131.

\bibitem[\protect\citeauthoryear{Garlappi and Skoulakis}{Garlappi and
  Skoulakis}{2009}]{Garlappi2009}
Garlappi, L. and G.~Skoulakis (2009).
\newblock Numerical solutions to dynamic portfolio problems: The case for value
  function iteration using {T}aylor approximation.
\newblock {\em Computational Economics\/}~{\em 33}, 193--207.

\bibitem[\protect\citeauthoryear{Harlow}{Harlow}{1991}]{Harlow1991}
Harlow, W.~V. (1991).
\newblock Asset allocation in a downside-risk framework.
\newblock {\em Financial Analysts Journal\/}~{\em 47\/}(5), 28--40.

\bibitem[\protect\citeauthoryear{Hata, Nagai, and Sheu}{Hata
  et~al.}{2010}]{Hata2010}
Hata, H., H.~Nagai, and S.-J. Sheu (2010).
\newblock Asymptotics of the probability minimizing a ``down-side'' risk.
\newblock {\em Annals of Applied Probability\/}~{\em 20\/}(1), 52--89.

\bibitem[\protect\citeauthoryear{Jain and Oosterlee}{Jain and
  Oosterlee}{2015}]{Jain2015}
Jain, S. and C.~W. Oosterlee (2015).
\newblock The stochastic grid bundling method: efficient pricing of {B}ermudan
  options and their {G}reeks.
\newblock {\em Applied Mathematics and Computation\/}~{\em 269\/}(1), 412--431.

\bibitem[\protect\citeauthoryear{Kharroubi, Langren\'e, and Pham}{Kharroubi
  et~al.}{2014}]{Kharroubi2014}
Kharroubi, I., N.~Langren\'e, and H.~Pham (2014).
\newblock A numerical algorithm for fully nonlinear {HJB} equations: an
  approach by control randomization.
\newblock {\em Monte Carlo Methods and Applications\/}~{\em 20\/}(2), 145--165.

\bibitem[\protect\citeauthoryear{Klebaner, Landsman, Makov, and Yao}{Klebaner
  et~al.}{2017}]{Klebaner2017}
Klebaner, F., Z.~Landsman, U.~Makov, and J.~Yao (2017).
\newblock Optimal portfolios with downside risk.
\newblock {\em Quantitative Finance\/}~{\em 17\/}(3), 315--325.

\bibitem[\protect\citeauthoryear{Konno, Shirakawa, and Yamazaki}{Konno
  et~al.}{1993}]{Konno1993}
Konno, H., H.~Shirakawa, and H.~Yamazaki (1993).
\newblock A mean-absolute deviation-skewness portfolio optimization model.
\newblock {\em Annals of Operations Reseach\/}~{\em 45\/}(1), 205--220.

\bibitem[\protect\citeauthoryear{Lai}{Lai}{1991}]{Lai1991}
Lai, T. (1991).
\newblock Portfolio selection with skewness: A multiple-objective approach.
\newblock {\em Review of Quantitative Finance and Accounting\/}~{\em 1\/}(3),
  293--305.

\bibitem[\protect\citeauthoryear{Longerstaey}{Longerstaey}{1996}]{Longerstaey1996}
Longerstaey, J. (1996).
\newblock Risk{M}etrics - technical document.
\newblock Technical report, J.P. Morgan.

\bibitem[\protect\citeauthoryear{Longstaff and Schwartz}{Longstaff and
  Schwartz}{2001}]{Longstaff2001}
Longstaff, F. and E.~Schwartz (2001).
\newblock Valuing {A}merican options by simulation: A simple least-squares
  approach.
\newblock {\em Review of Financial Studies\/}~{\em 14\/}(1), 681--692.

\bibitem[\protect\citeauthoryear{Markowitz}{Markowitz}{1952}]{Markowitz1952}
Markowitz, H. (1952).
\newblock Portfolio selection.
\newblock {\em Journal of Finance\/}~{\em 7\/}(1), 77--91.

\bibitem[\protect\citeauthoryear{Markowitz}{Markowitz}{1959}]{Markowitz1959}
Markowitz, H. (1959).
\newblock {\em Portfolio Selection: Efficient Diversification of Investment}.
\newblock New York: John Wiley and Sons.

\bibitem[\protect\citeauthoryear{Milevsky, Moore, and Young}{Milevsky
  et~al.}{2006}]{Milevsky2006}
Milevsky, M.~A., K.~S. Moore, and V.~R. Young (2006).
\newblock Asset allocation and annuity-purchase strategies to minimize the
  probability of financial ruin.
\newblock {\em Mathematical Finance\/}~{\em 16\/}(4), 647--671.

\bibitem[\protect\citeauthoryear{Morton, Popova, and Ivilina}{Morton
  et~al.}{2006}]{Morton2006}
Morton, D.~P., E.~Popova, and P.~Ivilina (2006).
\newblock Efficient fund of hedge funds construction under downside risk
  measures.
\newblock {\em Journal of Banking and Finance\/}~{\em 30\/}(2), 503--518.

\bibitem[\protect\citeauthoryear{Nagai}{Nagai}{2012}]{Nagai2012}
Nagai, H. (2012).
\newblock Downside risk minimization via a large deviation approach.
\newblock {\em Annals of Applied Probability\/}~{\em 22\/}(2), 608--669.

\bibitem[\protect\citeauthoryear{Pham}{Pham}{2003}]{Pham2003}
Pham, H. (2003).
\newblock A large deviations approach to optimal long term investment.
\newblock {\em Finance and Stochastics\/}~{\em 7\/}(2), 169--195.

\bibitem[\protect\citeauthoryear{Rockafellar and Uryasev}{Rockafellar and
  Uryasev}{2000}]{Rockafellar2000}
Rockafellar, R. and S.~Uryasev (2000).
\newblock Optimization of conditional value-at-risk.
\newblock {\em Journal of Risk\/}~{\em 2\/}(3), 21--42.

\bibitem[\protect\citeauthoryear{Tsitsiklis and Van~Roy}{Tsitsiklis and
  Van~Roy}{2001}]{Tsitsiklis01}
Tsitsiklis, J. and B.~Van~Roy (2001).
\newblock Regression methods for pricing complex {A}merican-style options.
\newblock {\em IEEE Transactions on Neural Networks\/}~{\em 12\/}(4), 694--703.

\bibitem[\protect\citeauthoryear{Tversky and Kahneman}{Tversky and
  Kahneman}{1992}]{Tversky1992}
Tversky, A. and D.~Kahneman (1992).
\newblock Advances in prospect theory: cumulative representation of
  uncertainty.
\newblock {\em Journal of Risk and Uncertainty\/}~{\em 5\/}(4), 297--323.

\bibitem[\protect\citeauthoryear{Van~Binsbergen and Brandt}{Van~Binsbergen and
  Brandt}{2007}]{vanBinsbergen2007}
Van~Binsbergen, J.~H. and M.~Brandt (2007).
\newblock Solving dynamic portfolio choice problems by recursing on optimized
  portfolio weights or on the value function?
\newblock {\em Computational Economics\/}~{\em 29}, 355--367.

\bibitem[\protect\citeauthoryear{von Neumann and Morgenstern}{von Neumann and
  Morgenstern}{1944}]{vonNeumann1944}
von Neumann, J. and O.~Morgenstern (1944).
\newblock {\em Theory of Games and Economic Behavior}.
\newblock Princeton University Press.

\bibitem[\protect\citeauthoryear{Winkelbauer}{Winkelbauer}{2014}]{Winkelbauer2014}
Winkelbauer, A. (2014).
\newblock Moments and absolute moments of the normal distribution.
\newblock arXiv:1209.4340.

\bibitem[\protect\citeauthoryear{Zhang, Langren\'e, Tian, Zhu, Klebaner, and
  Hamza}{Zhang et~al.}{2019}]{Zhang2018}
Zhang, R., N.~Langren\'e, Y.~Tian, Z.~Zhu, F.~Klebaner, and K.~Hamza (2019).
\newblock Dynamic portfolio optimization with liquidity cost and market impact:
  a simulation-and-regression approach.
\newblock {\em Quantitative Finance\/}~{\em 19\/}(3), 519--532.

\end{thebibliography}

\end{document}